\documentclass[10pt]{elsarticle}

\usepackage{bm}
\usepackage{mathtools}
\usepackage{graphicx}
\usepackage{multirow}
\newcommand{\norm}[1]{\left\lVert #1 \right\rVert}
\usepackage{algorithm}
\usepackage{algpseudocode}
\usepackage{algpascal}
 \usepackage{textcomp}

 \usepackage{tikz}
 \newcommand{\pkg}[1]{{\normalfont\fontseries{b}\selectfont #1}}
 \let\proglang=\textsf
 
 \usepackage{smartdiagram}
 \usepackage[utf8]{inputenc}
 \usetikzlibrary{arrows,positioning}
\usepackage{makecell} 
\setcellgapes{5pt}

\usepackage{amsmath,amsfonts,amssymb}

\DeclareMathOperator*{\argmin}{arg\,min}

\smartdiagramset{border color=none,
	set color list={blue!50!cyan,green!60!lime,orange!50!red,red!80!black},
	back arrow disabled=true, module minimum width = 4}

\smartdiagramset{module minimum height=1cm,
	module minimum width=0.8cm,
	module x sep=2}

\usepackage{hyperref}

\journal{Fuzzy Sets and Systems}

\begin{document}

\begin{frontmatter}

\title{Quantile-based fuzzy clustering of multivariate time series in the frequency domain}


\author[mymainaddress]{\'Angel L\'opez-Oriona\corref{mycorrespondingauthor} (ORCID 0000-0003-1456-7342)}
\ead{oriona38@hotmail.com}

\author[mymainaddress,mysecondaryaddress]{Jos\'e A. Vilar (ORCID 0000-0001-5494-171X)}
\cortext[mycorrespondingauthor]{Corresponding author}
\ead{jose.vilarf@udc.es}

\author[dursoaddress]{Pierpaolo D'Urso (ORCID 0000-0002-7406-6411)}
\ead{pierpaolo.durso@uniroma1.it}

\address[mymainaddress]{Research Group MODES, Research Center for Information and Communication Technologies (CITIC), University of A Coru\~na, 15071 A Coru\~na, Spain.}
\address[mysecondaryaddress]{Technological Institute for Industrial Mathematics (ITMATI), Spain.}
\address[dursoaddress]{Department of Social Sciences and Economics, Sapienza University of Rome, P. le Aldo Moro 5, Roma, Italy.}


\begin{abstract}
A novel procedure to perform fuzzy clustering of multivariate time series generated from different dependence models is proposed. Different amounts of dissimilarity between the generating models or changes on the dynamic behaviours over time are some arguments justifying a fuzzy approach, where each series is associated to all the clusters with specific membership levels. Our procedure considers quantile-based cross-spectral features and consists of three stages: (i) each element is characterized by a vector of proper estimates of the quantile cross-spectral densities, (ii) principal component analysis is carried out to capture the main differences reducing the effects of the noise, and (iii) the squared Euclidean distance between the first retained principal components is used to perform clustering through the standard fuzzy $C$-means and fuzzy $C$-medoids algorithms. The performance of the proposed approach is evaluated in a broad simulation study where several types of generating processes are considered, including linear, nonlinear and dynamic conditional correlation models. Assessment is done in two different ways: by directly measuring the quality of the resulting fuzzy partition and by taking into account the ability of the technique to determine the overlapping nature of series located equidistant from well-defined clusters. The procedure is compared with the few alternatives suggested in the literature, substantially outperforming all of them whatever the underlying process and the evaluation scheme. Two specific applications involving air quality and financial databases illustrate the usefulness of our approach. 

	
\end{abstract}


\end{frontmatter}


\section{Introduction}

Time series clustering is a pivotal problem in data mining with applications in a wide variety of fields, including machine learning, economics, finance, physics, computer science, medicine, biology, geology, among others. The goal is to split a large set of unlabeled time series realizations into homogeneous groups so that similar series are placed together in the same group and dissimilar series are located in different groups. This unsupervised classification process is useful to detect different dynamic patterns without the need to analyse and model each single time series, which is computationally intensive and often far from being the real target. Many methods to cluster time series have been proposed in the literature during the last two decades. Comprehensive overviews including current advances, future prospects, significant references and specific application areas are provided by \cite{Liao:2005, Fu:2011, Rani_Sikka:2012, aghabozorgi2015time}, and more recently in the monograph by \cite{MaharajDUrsoCaiado:2019}. However, most of the proposed approaches concern univariate time series (UTS) while clustering of multivariate time series (MTS) has received much less attention. Unlike UTS, MTS involve a number of variables which must be jointly considered to characterize the underlying dynamic pattern. From the grouping point of view, this is a challenging issue because a dissimilarity measure between MTS should take into account the interdependence relationship between variables. For example, the cross-correlation between some specific dimensions might be high in some clusters but non-significant in others. Additionally, MTS are two-dimensional objects, which increases the computational complexity, making inefficient or even infeasible some of the clustering procedures proposed to deal with UTS. In short, high dimensionality and complexity to assess dissimilarity make particularly challenging the MTS clustering task.

There exist two important decisions to be made in any clustering problem, namely the notion of dissimilarity between the elements subject to the grouping problem and the clustering algorithm determining how the assignment of objects to the groups is done. A proper dissimilarity measure must be totally dependent on the nature and specific purpose of the clustering task, thus providing a meaningful clustering solution according to the grouping target. If the goal is to discriminate between geometric profiles of the time series, then a shape-based dissimilarity criterion is suitable. In contrast, a structure-based dissimilarity is desirable if the intention is to compare underlying dependence models. In the latter case, the clustering performance may be seriously affected by noise, change in the conditional variance or heavy-tailed errors, and hence distance measures capable of capturing high level dynamic structures are particularly helpful. Many criteria to assess dissimilarity between UTS are available in the literature, including measures based on raw data, extracted features, generating models, complexity levels, and forecast performances, among others. A survey of measures can be seen in \cite{MonteroVilar:2014} and many of them are implemented in the \proglang{R} package \pkg{TSclust} \cite{MonteroVilarTSclust:2014}.

Regarding how the assignment of the elements to the clusters is made, there are two classical paradigms which are usually referred to as ``hard'' and ``soft'' clustering. The partition provided by hard clustering procedures locates each data object in exactly one cluster, thus being constituted by disjoint subsets. This approach does not allow for overlapping clusters and could become too inflexible in some scenarios. For instance, hard clustering is incapable of giving insights into which elements are equidistant from two or more clusters and also to account for a closer alignment to patterns of other clusters due to changes in the dynamic of the series over time. On the other hand, fuzzy clustering strategies \cite{bezdek2013pattern, miyamoto2008algorithms} provide a more versatile approach to address the clustering task. They rely on the notion of membership of an element in a given cluster, which indicates the degree of confidence in that particular assignment. Therefore, the output of these methods is a soft partition where the objects can belong to several groups with specific membership degrees. 

Several works have considered fuzzy clustering of time series, specially in the univariate setting. \cite{d2009autocorrelation} proposed a fuzzy clustering approach based on estimates of the autocorrelation function of the time series up to a given lag. The corresponding quantities are used as input to the traditional fuzzy $C$-means algorithm. \cite{maharaj2011fuzzy} provided three different methods relying on different characteristics computed from the UTS, namely the periodogram, the normalized periodogram and the logarithm of the normalized periodogram. A fuzzy approach to the clustering of UTS based on estimated wavelet variances is presented in \cite{maharaj2010wavelet}, proving itself capable of identifying time series with switching patterns in terms of variability. \cite{d2017fuzzy} developed an approach focused on grouping together time series with similar seasonal structures using extreme value analysis. The input features in the fuzzy clustering algorithm are parameter estimates of time varying location, scale and shape obtained by means of a fitting of the generalised extreme value distribution. \cite{d2013clustering} developed two fuzzy clustering strategies aimed at clustering financial time series. The first approach employs the autoregressive representation of GARCH models whereas the second one is based on estimates of GARCH parameters. A method for grouping heteroskedastic time series was designed by \cite{d2016garch}. The approach assumes that the series follow a GARCH model. Estimates of the so-called unconditional volatility and time-varying volatility are obtained and then used to feed the classical fuzzy $C$-medoids technique. Robust alternatives of the method are also provided. \cite{vilar2018quantile} extended their work in \cite{lafuente2016clustering} by introducing a procedure which employs estimates of the quantile autocovariance function in order to perfom fuzzy clustering of UTS by means of the fuzzy $C$-medoids algorithm. The method takes advantage of the nice properties of the quantile autocovariance function as robustness to heavy tails or no requirements about the existence of moments. Robust approaches derived from this technique are developed in \cite{lafuente2018robust}. \cite{disegna2017copula} introduced a copula-based procedure for finding groups in spatial time series. The methodology obtains the empirical copula for a given MTS and computes a dissimilarity between this copula end the Frechet copula by taking into account also spatial information. Note that all the previous mentioned approaches pertain to the so-called feature-based clustering, which exploits specific features extracted from a time series, or to the termed model-based clustering, which groups the time series based on estimates of the parameters or of the residuals of a given model. 

 Approaches based on direct computation of a given distance have also been suggested for fuzzy clustering of UTS. \cite{izakian2015fuzzy} designed an approach employing the dynamic time warping distance (DTW) in the standard fuzzy $C$-means algorithm. A specific procedure to obtain the average of a set of UTS according to DTW is utilised. A trimmed fuzzy clustering technique for financial time series relying on DTW and the fuzzy $C$-medoids algorithm is developed in \cite{d2019trimmed}. The method is designed so as to attain robustness against outlying elements by trimming away the series which are more distant from the bulk of the data. \cite{moller2003fuzzy} proposed a fuzzy clustering procedure for short time series that is motivated by experiments in molecular biology. The fuzzy $C$-means algorithm is used in combination with the termed short time series distance, which is capable of measuring both shape-based and time-based similarities.

 By contrast, a fewer number of papers have dealt with fuzzy clustering of MTS. \cite{d2005fuzzy} introduced three fuzzy $C$-means clustering approaches of multivariate time trajectories considering the so-called positional dissimilarity, velocity dissimilarity and a mixture of both. These models are particularly beneficial when anomalous trajectories are present in the dataset. \cite{coppi2006fuzzy} proposed to perform unsupervised classification of MTS by means of a novel objective function containing two terms, one taking into account longitudinal features of the MTS and the other considering the Shannon entropy measure concerning fuzzy partitions. A procedure based on the maximum overlap discrete wavelet transform is provided in \cite{d2012wavelets}. Estimates of wavelet variances and correlations are computed from each MTS and used to feed the fuzzy $C$-means, fuzzy $C$-medoids and fuzzy relational clustering algorithms. \cite{he2018unsupervised} devised an approach where each MTS sample is treated as matrix data. First, a dimensionality reduction technique is applied over the original MTS dataset, and then a fuzzy clustering algorithm based on spatial weighted distance matrix is performed on the reduced dataset. Four different robust clustering models for MTS aimed at neutralizing the negative effects of outliers are developed in \cite{d2018robust}. All of them take into account the exponential transformation, which assigns ``small weights'' to outliers in the clustering process, hence achieving the desired robustness. \cite{li2020fuzzy} designed a sophisticated feature-weighted clustering method based on two dissimilarity measures, DTW and the named shape-based distance. The approach consists of several steps in which the contribution of each independent dimension to the overall clustering process is considered. Note that, except for \cite{d2012wavelets}, all the previously cited works on fuzzy clustering of MTS are not suitable for making the grouping in terms of underlying dependence structures, since they are aimed to measure dissimilarity in shape. Thus, it is clear that there is a need for developing fuzzy clustering approaches capable of addressing the task from the former perspective.

 This paper is aimed at evaluating the performance of a distance measure based on the quantile cross-spectral density (QCD) in fuzzy clustering of MTS. Our assumption is that the goal consists of grouping the series in terms of their generating processes, that is, we consider two MTS to be similar if their underlying dependence patterns are alike. Note that this premise is quite realistic if, for instance, one intends to detect the temporal pattern driving a time series which is observed in practice under large amounts of noise. Under this criterion, a metric capable of effectively discriminating between different generating mechanisms while displaying a large degree of robustness to the type of underlying processes is undoubtedly needed in order to attain a meaningful clustering solution. Indeed, QCD always exists under the assumption of strictly stationarity and allows to account for complex forms of dependence that other time series features as traditional autocovariances are unable to uncover \cite{Barunik_Kley:2019}. In addition, QCD takes advantage of automatically considering all the lags. These nice properties insinuate that a dissimilarity measure based on QCD can achieve great results in grouping MTS according to the stated goal. In our previous work \cite{oriona2020}, we decided to answer this question by analysing the performance of a distance measure termed $d_{QCD}$ in different scenarios of MTS clustering through a comprehensive simulation study. The results showed that $d_{QCD}$ is highly competitive when grouping linear processes and significantly outperforms alternative dissimilarities suggested in the literature when dealing with processes showing a high amount of heteroskedascity or complicated types of dependence. Moreover, $d_{QCD}$ exhibited a substantial degree of robustness against changes in the distributional form of the error terms.
 
 The main contribution of this work consists of proposing a novel fuzzy procedure for MTS clustering taking advantage of the high capability of the quantile cross-spectral density to characterize any type of serial dependence structure. Specifically, we take into consideration a modified version of the distance $d_{QCD}$ given in \cite{oriona2020}. While $d_{QCD}$ is directly constructed from estimates of QCD, the alternative metric we propose here considers the principal component analysis (PCA) transformation over the mentioned estimates. This way, a lot of the noise contained in these (correlated) estimates is removed and the most important information is retained, thus getting improved performance in comparison with the original distance while inheriting all its advantageous characteristics. The new distance is utilised in both the standard fuzzy $C$-means and the fuzzy $C$-medoids algorithms. Assessment of the proposed approaches is carried out by means of an extensive simulation study including linear, nonlinear and conditional heteroskedastic processes. The dissimilarity proposed in \cite{d2012wavelets} and a natural extension of that introduced in \cite{d2009autocorrelation} are also analysed for comparison purposes. Two evaluation schemes are considered. The first one is aimed at examining the capability of the procedures in assigning high (low) membership values if a given series pertains (not pertains) to a specific cluster defined in advance. The second scheme also analyses the ability of the approaches to handle outlying series. Lastly, two specific applications to multivariate financial and environmental datasets are presented to highlight the usefulness of the proposed clustering technique. 
 
 The remainder of this paper is structured as follows. Section \ref{sectionqcd} presents $d_{QCD}$, a dissimilarity measure between a pair of MTS that takes into consideration proper estimates of QCD. The estimation procedure is detailed and some powerful properties of the metric holding under very general conditions are highlighted. In Section \ref{sectionqcdpca}, the dissimilarity is considered to develop two novel fuzzy clustering approaches relying on the traditional fuzzy $C$-means and fuzzy $C$-medoids algorithms. A direct modification of the distance $d_{QCD}$ based on the PCA transformation is proposed. The new measure considers the transformed QCD-based features concerning the principal components space. Advantages of this metric in comparison with the original dissimilarity are shown by means of a toy example. The dissimilarity based on QCD and PCA is used to perform fuzzy clustering in Section \ref{sectionsimulationstudy}. Three scenarios characterised by the kind of generating process are considered, namely linear, nonlinear and dynamic conditional correlation. The assessment task is performed in a fair and general manner, and the results are compared with those obtained using alternative dissimilarity measures. Section \ref{sectionapplication} contains applications to real datasets and some concluding remarks are summarized in Section \ref{sectionconcludingremarks}. 

\section{A distance measure based on the quantile cross-spectral density}\label{sectionqcd}

Consider a set of $s$ multivariate time series $\mathcal{S} = \left\{ \bm{X}_t^{(1)}, \ldots, \bm{X}_t^{(s)} \right\}$, where the $j$-th element $\bm{X}_t^{(j)}= \left\{ \bm{X}_{1}^{(j)},\ldots ,\bm{X}_{T_j}^{(j)}\right\}$ is a $T_j$-length partial realization from any $d$-variate real-valued strictly stationary stochastic process $(\bm{X}_t)_{t\in\mathbb{Z}}$. We wish to perform clustering on the elements of $\mathcal{S}$ in such a way that the series generated from the same stochastic process are grouped together. We propose to use a partitional algorithm starting from a pairwise dissimilarity matrix based on comparing estimated quantile cross-spectral densities. In this section, the quantile cross-spectral density notion is presented and then used to define a distance between MTS.

\subsection{The quantile cross-spectral density}\label{subsectionqcd}

Let  $\{\bm{X}_t, \, t\in\mathbb{Z}\} = \{(X_{t,1},\ldots,X_{t,d}), \, t\in\mathbb{Z}\}$ be a $d$-variate real-valued strictly stationary stochastic process. Denote by $F_j$ the marginal distribution function of $X_{t,j}$, $j=1,\ldots,d$, and by $q_j(\tau)=F_j^{-1}(\tau)$, $\tau \in [0,1]$, the corresponding quantile function. Fixed $l \in \mathbb{Z}$ and an arbitrary couple of quantile levels $(\tau,\tau^{\prime}) \in [0,1]^2$, consider the cross-covariance of the indicator functions $ I\left\{ X_{t,j_1} \leq  q_{j_1}(\tau) \right\}$ and $I\left\{ X_{t+l, j_2} \leq  q_{j_2} (\tau^{\prime}) \right\}$ given by
\begin{equation}		\label{qac}
\gamma_{j_1,j_2}(l,\tau,\tau^{\prime}) = \mbox{Cov} 
\left(  I \left\{ X_{t, j_1} \leq q_{j_1}(\tau) \right\}, I \left\{ X_{t+l, j_2} \leq q_{j_2}(\tau^{\prime}) \right\} \right),
\end{equation}
for $1\leq j_1,j_2 \leq d$. Taking $j_1=j_2=j$, the function $\gamma_{j,j}(l,\tau,\tau^{\prime})$, with $(\tau,\tau^{\prime}) \in [0,1]^2$, so-called QAF of lag $l$, generalizes the traditional autocovariance function. While autocovariances measure linear dependence between different lags evaluating covariability with respect to the average, quantile autocovariances examine how a part of the range of variation of $X_{j}$ helps to predict whether the series will be below quantiles in a future time. This way, QAF entirely describes  the joint distribution of $( X_{t,j}, X_{t+l,j} )$, enabling us to capture serial features that standard autocovariances cannot detect. Note that $\gamma_{j_1,j_2}(l,\tau,\tau^{\prime})$ always exists since no assumptions about moments are required. Furthermore, QAF also takes advantage of the local distributional properties inherent to the quantile methods, including robustness against heavy tails, dependence in the extremes and changes in the conditional shapes (skewness, kurtosis). Motivated by these nice properties, a dissimilarity between UTS based on comparing estimated quantile autocovariances over a common range of quantiles was proposed by \cite{lafuente2016clustering} to perform UTS clustering with very satisfactory results. 

In the case of the multivariate process $\{\bm{X}_t, \, t\in\mathbb{Z}\}$, we can consider the $d \times d$ matrix
\begin{equation}  \label{gammaMTS}
\bm{\Gamma} (l,\tau,\tau^{\prime}) = 
\left( \gamma_{j_1,j_2}(l,\tau,\tau^{\prime}) \right)_{1\leq j_1,j_2\leq d},
\end{equation}
which jointly provides information about both the cross-dependence (when $j_1\neq j_2$) and the serial dependence (because the lag $l$ is considered). To obtain a much richer picture of the underlying dependence structure, $\bm{\Gamma} (l,\tau,\tau^{\prime})$ can be computed over a range of prefixed values of $L$ lags, $\mathcal{L}=\{l_1,\ldots,l_L\}$, and $r$ quantile levels, $\mathcal{T} = \{\tau_1,\ldots,\tau_r\}$, thus having available the set of matrices
\begin{equation}  \label{gammaProcess}
\bm{\Gamma}_{\bm{X}_t} \left(\mathcal{L},\mathcal{T}\right) = 
\left\{ \bm{\Gamma} (l,\tau,\tau^{\prime}), \,\,  l \in \mathcal{L}, \, \, \tau,\tau^{\prime} \in \mathcal{T} 
\right\}. 
\end{equation}

In the same way as the spectral density is the representation in the frequency domain of the autocovariance function, the spectral counterpart for the cross-covariances $\gamma_{j_1,j_2}(l,\tau,\tau^{\prime})$ can be introduced. Under suitable summability conditions (mixing conditions), the Fourier transform of the cross-covariances is well-defined and  the \textit{quantile cross-spectral density} is given by 
\begin{equation} \label{qcrossdens.j1j2}
{\mathfrak f}_{j_1,j_2} (\omega, \tau,\tau^{\prime}) = (1/2\pi) \sum_{l=-\infty}^{\infty} 
\gamma_{j_1,j_2}(l,\tau,\tau^{\prime}) e^{-il\omega},
\end{equation}
for $1\leq j_1,j_2 \leq d$, $\omega \in \mathbb{R}$ and $\tau,\tau^{\prime} \in [0,1]$. Note that ${\mathfrak f}_{j_1,j_2}(\omega, \tau,\tau^{\prime})$ is complex-valued so that it can be represented in terms of its real and imaginary parts, which will be denoted by $\Re({\mathfrak f}_{j_1,j_2}(\omega, \tau,\tau^{\prime}))$ and $\Im({\mathfrak f}_{j_1,j_2}(\omega, \tau,\tau^{\prime}))$, respectively. The quantity $\Re({\mathfrak f}_{j_1,j_2}(\omega, \tau,\tau^{\prime}))$ is known as quantile cospectrum of $(X_{t, j_1})_{t\in\mathbb{Z}}$ and $(X_{t, j_2})_{t\in\mathbb{Z}}$, whereas the quantity -$\Im({\mathfrak f}_{j_1,j_2}(\omega, \tau,\tau^{\prime}))$ is called quantile quadrature spectrum of $(X_{t, j_1})_{t\in\mathbb{Z}}$ and $(X_{t, j_2})_{t\in\mathbb{Z}}$. 

For fixed quantile levels $(\tau,\tau^{\prime})$, the quantile cross-spectral density is the cross-spectral density of the bivariate process

\begin{equation}
( I \{ X_{t, j_1} \leq q_{j_1}(\tau) \}, I \{ X_{t+l, j_2} \leq q_{j_2}(\tau^{\prime}) \} ).
\end{equation}

Therefore the quantile cross-spectral density measures dependence between two components of the multivariate process in different ranges of their joint distribution and across frequencies. Proceeding as in (\ref{gammaProcess}), the quantile cross-spectral density can be evaluated on a range of frequencies $\Omega$ and of quantile levels $\mathcal{T}$ for every couple of components in order to obtain a complete representation of the process, i.e., consider the set of matrices
\begin{equation}  \label{qcrossdens.X}
\bm{\mathfrak f}_{\bm{X}_t} \left(\Omega, \mathcal{T}\right) = 
\left\{ \bm{\mathfrak f} (\omega,\tau,\tau^{\prime}), \, \, \omega \in \Omega, \, \, \tau,\tau^{\prime} \in \mathcal{T} 
\right\}, 
\end{equation}
where $\bm{\mathfrak f} (\omega,\tau,\tau^{\prime})$ denotes the $d\times d$ matrix in $\mathbb{C}$
\begin{equation}   \label{qcrossdens}
\bm{\mathfrak f} (\omega,\tau,\tau^{\prime}) = \left( {\mathfrak f}_{j_1,j_2} (\omega, \tau,\tau^{\prime}) \right)_{1 \leq j_1,j_2 \leq d}.
\end{equation}

Representing $\bm{X}_t$ through $\bm{\mathfrak f}_{\bm{X}_t}$, a complete information on the general dependence structure of the process is available. Comprehensive discussions about the nice properties of the quantile cross-spectral density are given in \cite{Lee_SubbaRao:2012}, \cite{Dette_Hallin_Kley_Volgushev:2015} and \cite{Barunik_Kley:2019}, including invariance to monotone transformations, robustness and  capability to detect nonlinear dependence. It is also worth enhancing that the quantile cross-spectral density provides a full description of all copulas of pairs of components in $\bm{X}_t$, since the difference between the copula of an arbitrary couple $(X_{t,j_1},X_{t+l,j_2})$ evaluated in $(\tau,\tau^{\prime})$ and the independence copula at $(\tau,\tau^{\prime})$ can be written as 

\begin{equation}\label{copulaqcd}
 \mathbb{P} \left(  X_{t,j_1} \leq q_{j_1}(\tau), X_{t+l,j_2} \leq q_{j_2}(\tau^{\prime}) \right) - \tau \tau^{\prime}
= \int_{-\pi}^{\pi} {\mathfrak f}_{j_1,j_2} (\omega, \tau,\tau^{\prime}) e^{il\omega} \, d \omega.
\end{equation}

According with the prior arguments, a dissimilarity measure between realizations of two multivariate processes, $\bm{X}_t$ and $\bm{Y}_t$, could be established by comparing their representations in terms of the quantile cross-spectral density matrices, $\bm{\mathfrak f}_{\bm{X}_t}$ and $\bm{\mathfrak f}_{\bm{Y}_t}$, respectively. For it, estimates of the quantile cross-spectral densities must be obtained. 

Let $\left\{ \bm{X}_{1},\ldots ,\bm{X}_{T} \right\}$ be a realization from the process $(\bm{X}_t)_{t\in\mathbb{Z}}$ so that $\bm{X}_{t}=(X_{t,1},\ldots,X_{t,d})$, $t=1,\ldots,T$. For arbitrary $j_1, j_2 \in \{1,\ldots,d\}$ and $(\tau,\tau^{\prime}) \in [0,1]^2$, \cite{Barunik_Kley:2019} propose to estimate ${\mathfrak f}_{j_1,j_2} (\omega, \tau,\tau^{\prime})$ considering a smoother of the cross-periodograms based on the indicator functions $I\{ \hat{F}_{T,j}(X_{t,j})\}$, where $\hat{F}_{T,j} (x) = T^{-1} \sum_{t=1}^{T} I\{ X_{t,j} \leq  x\}$ denotes the empirical distribution function of $X_{t,j}$. This approach extends to the multivariate case the estimator proposed by \cite{Kley_Volgushev_Dette_Hallin:2016} in the univariate setting. More specifically, the called \textit{rank-based copula cross periodogram} (CCR-periodogram) is defined by
\begin{equation}   \label{CCR}
I^{j_1,j_2}_{T,R} (\omega,\tau,\tau^{\prime}) = \frac{1}{2 \pi T} d^{j_1}_{T,R} (\omega,\tau) 
d^{j_2}_{T,R} (-\omega,\tau^{\prime}),
\end{equation}
where 
$$d_{T,R}^{j} (\omega,\tau) = \sum_{t=1}^{T} I\{ \hat{F}_{T,j}(X_{t,j}) \leq \tau \} e^{-i\omega t}.$$
The asymptotic properties of the CCR-periodogram are established in Proposition 4.1 of \cite{Barunik_Kley:2019}. Likewise the standard cross-periodogram, the CCR-periodogram is not a consistent estimator of ${\mathfrak f}_{j_1,j_2} (\omega, \tau,\tau^{\prime})$ \cite{Kley_Volgushev_Dette_Hallin:2016}. To achieve consistency, the CCR-periodogram ordinates (evaluated on the Fourier frequencies) are convolved with weighting functions $W_T(\cdot)$. The \textit{smoothed CCR-periodogram} takes the form
\begin{equation}   \label{smoothedCCR}
\hat{G}^{j_1,j_2}_{T,R} (\omega,\tau,\tau^{\prime}) = (2 \pi/T) \sum_{s=1}^{T-1} W_T \left( \omega- \frac{2 \pi s}{T} \right)  
I^{j_1,j_2}_{T,R} \left( \frac{2 \pi s}{T} ,\tau,\tau^{\prime} \right),
\end{equation}
where 
$$W_T \left( u \right) = \sum_{v=-\infty}^{\infty} (1/h_T) W\left( 
\frac{u + 2 \pi v}{h_T} \right),  $$
with $h_T>0$ a sequence of bandwidths such that $h_T \rightarrow 0$ and $Th_T \rightarrow \infty$ as $T\rightarrow \infty$, and $W$ is a real-valued, even weight function with support $[-\pi, \pi]$. 
Consistency and asymptotic performance of the smoothed CCR-periodogram $\hat{G}^{j_1,j_2}_{T,R} (\omega,\tau,\tau^{\prime})$ are established in Theorem 3.5 of \cite{Kley_Volgushev_Dette_Hallin:2016}. 

This way, the set of complex-valued matrices $\bm{\mathfrak f}_{\bm{X}_t} \left(\Omega, \mathcal{T}\right)$ in (\ref{qcrossdens.X}) characterizing the underlying process can be estimated by 
\begin{equation}  \label{est.qcrossdens.X}
\hat{\bm{\mathfrak f}}_{\bm{X}_t} \left(\Omega, \mathcal{T}\right) = 
\left\{ \hat{\bm{\mathfrak f}} (\omega,\tau,\tau^{\prime}), \, \, \omega \in \Omega, \, \, \tau,\tau^{\prime} \in \mathcal{T} 
\right\}, 
\end{equation}
where $\hat{\bm{\mathfrak f}} (\omega,\tau,\tau^{\prime})$ is the matrix
\begin{equation}   \label{est.qcrossdens}
\hat{\bm{\mathfrak f}} (\omega,\tau,\tau^{\prime}) = \left( 
\hat{G}^{j_1,j_2}_{T,R} (\omega,\tau,\tau^{\prime}) 
\right)_{1 \leq j_1,j_2 \leq d}.
\end{equation}

Throughout this article, the smoothed CCR-periodograms were obtained by using the \proglang{R}-package \pkg{quantspec} \cite{Kley_quantspec}.

\subsection{A spectral dissimilarity measure between MTS}\label{subsectiondqcd}

A simple dissimilarity criterion between a pair of $d$-variate time series $\bm X_t^{(1)}$ and $\bm X_t^{(2)}$ can be obtained by comparing their estimated sets of complex-valued matrices $\hat{\bm{\mathfrak f}}_{\bm{X}_t^{(1)}} \left(\Omega, \mathcal{T}\right)$ and $\hat{\bm{\mathfrak f}}_{\bm{X}_t^{(2)}} \left(\Omega, \mathcal{T}\right)$ evaluated on a common range of frequencies and quantile levels. Specifically, each time series $\bm X_t^{(u)}$, $u=1,2$, is characterized by means of a set of $d^2$ vectors \{$\bm \Psi_{j_1,j_2}^{(u)},1\leq j_1, j_2 \leq d \}$ constructed as follows. For a given set of $K$ different frequencies $\Omega = \{\omega_1, \ldots, \omega_K\}$, and $r$ quantile levels $\mathcal{T}=\{\tau_1,\ldots,\tau_r\}$, each vector $\bm \Psi_{j_1,j_2}^{(u)}$ is given by


\begin{equation}\label{uj1j2}
\bm \Psi_{j_1,j_2}^{(u)}=(\bm \Psi_{1,j_1,j_2}^{(u)} , \ldots, \bm \Psi_{K,j_1,j_2}^{(u)}),
\end{equation}

\noindent where each $\bm \Psi_{k,j_1,j_2}^{(u)}$, $k=1,\ldots, K$, consists of a vector of length $r^2$ formed by rearranging by rows the elements of the matrix

\begin{equation}
(\hat{\bm{\mathfrak f}} (\omega_k,\tau_i,\tau_{i^{\prime}}); i,i^\prime=1,\ldots, r),
\end{equation}

\noindent with $\hat{\bm{\mathfrak f}} (\omega_k,\tau_i,\tau_{i^{\prime}}) \in \hat{\bm{\mathfrak f}}_{\bm{X}_t^{(u)}} \left(\Omega, \mathcal{T}\right)$.

Once the set of $d^2$ vectors $\bm \Psi_{j_1,j_2}^{(u)}$ is obtained, they are all concatenated in a vector $\bm \Psi^{(u)}$ in the same way as vectors $\bm \Psi_{k,j_1,j_2}^{(u)}$ constitute $\bm \Psi_{j_1,j_2}^{(u)}$ in \eqref{uj1j2}. In this manner, the dissimilarity between $\bm X_t^{(1)}$ and $\bm X_t^{(2)}$ is obtained by means of the Euclidean distance between $\bm \Psi^{(1)}$ and $\bm \Psi^{(2)}$
\begin{equation}  \label{d.QCD}
\begin{split}
d_{QCD}(\bm X_t^{(1)}, \bm X_t^{(2)})= \Big[||\Re_v(\bm \Psi^{(1)})-\Re_v(\bm \Psi^{(2)})||^2+||\Im_v(\bm \Psi^{(1)})-\Im_v(\bm \Psi^{(2)})||^2\Big]^{1/2}=\\
\Bigg[
\sum_{j_1=1}^{d}\sum_{j_2=1}^{d}\sum_{i=1}^{r}\sum_{i'=1}^{r}\sum_{k=1}^{K}\Big(\Re({\hat G^{j_1,j_2}_{T,R}(\omega_{k}, \tau_{i}, \tau_{i^ {\prime}})^{(1)}})-\Re({\hat G^{j_1,j_2}_{T,R}(\omega_{k}, \tau_{i}, \tau_{i^ {\prime}})^{(2)})}\Big)^2+\\
\sum_{j_1=1}^{d}\sum_{j_2=1}^{d}\sum_{i=1}^{r}\sum_{i'=1}^{r}\sum_{k=1}^{K}\Big(\Im({\hat G^{j_1,j_2}_{T,R}(\omega_{k}, \tau_{i}, \tau_{i^ {\prime}})^{(1)}})-\Im({\hat G^{j_1,j_2}_{T,R}(\omega_{k}, \tau_{i}, \tau_{i^ {\prime}})^{(2)})}\Big)^2
\Bigg]^{1/2},
\end{split}
\end{equation}

\noindent where $\Re_v$ and $\Im_v$ denote the element-wise real and imaginary part operations, respectively, and $\hat G^{j_1,j_2}_{T,R}(\omega_{k}, \tau_{i}, \tau_{i^ {\prime}})^{(u)}$ is the corresponding element of the matrix given by \eqref{est.qcrossdens} for the series $\bm X_t^{(u)}$.





Computation of vectors $\bm \Psi^{(1)}, \ldots, \bm \Psi^{(n)}$ for every MTS in the set $S$ could be used to perform fuzzy clustering in $S$ by means of an algorithm as fuzzy $C$-means or fuzzy $C$-medoids considering the distance $d_{QCD}$. This distance has been successfully applied to perform clustering on MTS in a crisp framework \cite{oriona2020}, and the corresponding QCD-based features, to develop classification \cite{oriona2021a} and outlier detection \cite{oriona2021b} procedures.

\subsection{Properties of $d_{QCD}$}\label{subsectionpropertiesdqcd}

Now we present some important properties of the distance $d_{QCD}$ which make it a very powerful dissimilarity to perform clustering of MTS. 

In the following, we assume that $\bm X_t^{i}$ is a $d$-variate, real-valued, strictly stationary process and $\bm X_t^{(i)}$ is a realization of length $T$ from the process $\bm X_t^{i}$. The $j$-th component of $\bm X_t^{i}$, $j=1,\ldots,d$, is denoted by $X_{t,j}^{i}$. The notation $F_j^i$ stands for the marginal cumulative distribution function of $X_{t,j}^i$. Given a lag $l \in \mathbb{Z}$ and a couple of components $j_1,j_2 = 1,\ldots, d$, the joint cumulative distribution function of the pair $(X_{t,j_1}^{i}, X_{t+l,j_2}^{i})$ is denoted by $F_{j_1, j_2,l}^i$. We suppose that all the mentioned cumulative distribution functions are continuous functions. Now we state the following properties. \\ \\
\noindent \textbf{Property 1}. If $ \bm X_t^1=\bm X_t^2$. Then $d_{QCD}(\bm{X}_t^{(1)}, \bm{X}_t^{(2)}) \xrightarrow[p]{}0$ as $T \xrightarrow{}{}\infty$, where the notation $\xrightarrow[p]{}$ stands for convergence in probability. \\ 

\noindent \textbf{Property 2}. Assume that there exists some $l \in \mathbb{Z}$ and a couple of dimensions $j_1,j_2 = 1,\ldots, d$ such that $F_{j_1,j_2,l}^1\ne F_{j_1,j_2,l}^2$ and that $F_j^1=F_{j}^2$, $j=1, \ldots, d$. Then there exist an infinite number of probability levels and an infinite number of frequencies such that $d_{QCD}(\bm{X}_t^{(1)}, \bm{X}_t^{(2)})\xrightarrow[p]{}a$, $a \ne 0$ as $T \xrightarrow{}{}\infty$. 

The proofs of the previous properties are deferred to the appendix so as not to impair the flow of the paper. Some remarks about the results are given below. \\

\noindent \textbf{Remark 1}. Property 1 is a desirable characteristic of any distance measure aimed at performing clustering of MTS based on underlying dependence patterns. Indeed, it tells us that the distance between two MTS generated from the same process is expected to be negligible for a sufficiently large value of the series length. The majority of the distance measures suggested in the literature for UTS or MTS clustering according to the stated goal have this property. \\

\noindent \textbf{Remark 2}: Property 2 assumes that the marginal distribution function of the $j$-th component of $\bm X_t^{1}$ is equal to the marginal distribution function of the $j$-th component of $\bm X_t^{2}$. This can be supposed without loss of generality. Indeed, if $F_{j'}^1 \ne F_{j'}^2$ for some $j' \in \{1, \ldots, d\}$, then the quantile autocovariance in \eqref{qac} regarding $l=0$ and $j_1=j_2=j'$ is going to be different between both processes for some pair of probability levels $(\tau,\tau^{\prime}) \in [0,1]^2$. It is not difficult to derive (see \eqref{qcrossdens.j1j2}) that this discrepancy is transmitted to the corresponding smoothed CCR-periodograms, resulting in the convergence in probability of the respective distance between two realizations of the processes to a quantity distinct from zero. \\

\noindent \textbf{Remark 3}: Property 2 can be directly extended to more than two processes. In fact, given a collection of $N$ processes $\bm X_t^{b}$ and the corresponding realizations $\bm X_t^{(b)}$, $b=1, \ldots, N$, if we assume that (1) for every pair $(i, i')$, $i, i' \in \{1, \ldots, N\}$, there exist $j_{i}$, $j_{i'}$ and $l_{i, i'} \in \mathbb{Z}$ such that $F_{j_i,j_{i'}, l_{i, i'}}^i\ne F_{j_i,j_{i'}, l_{i, i'}}^{i'}$, (2) every couple of functions of the form $F_{j_k,j_{k'}, l_{k, k'}}^k$, $k, k' \in  \{1, \ldots, N\}$ are different from one another, and (3) $F_j^1=F_j^2=\ldots =F_j^N$, $j=1,\ldots,d$, then there exist an infinite number of probability levels, an infinite number of frequencies, and a set of different real numbers $\{a_{q, q'}\}$, $q, q' = 1, \ldots, N$, $q \ne q'$ such that $d_{QCD}(\bm{X}_t^{(q)}, \bm{X}_t^{(q')})\xrightarrow[p]{}$ $a_{q,q'} \ne 0$ as $T \xrightarrow{}{}\infty$. Note that this is often the case in practice when we apply cluster analysis through $d_{QCD}$ to a set of MTS coming from more than two different generating processes. \\

\noindent \textbf{Remark 4}: Property 2 is perhaps the most important characteristic of the distance $d_{QCD}$. Broadly speaking, it tells us that, under appropriate conditions, the metric is able to capture even the slightest change in the dependence structure between two generating processes. This trait is not shared, to the best of our knowledge, by any of the metrics suggest in the literature for clustering of UTS or MTS based in dependence patterns. Generally, it is easy to find a counterexample where two MTS are produced from distinct generating processes but the corresponding dissimilarities fail to detect any difference between the corresponding realizations even for very large values of the series lengths. In fact, our previous work \cite{oriona2020} provides an insightful example of this happening. There we simulated bivariate MTS from three types of the so-called QVAR processes. These kind of processes are capable of generating rich forms of quantile dependence while keeping uncorrelatedness within and between components. The results shown in that work determine that most of the alternative metrics were totally unable to unmask the underlying patterns (see the middle panel in Figure 3 of \cite{oriona2020}), whereas the distance $d_{QCD}$ perfectly discriminated between processes. This property is due to the relationship between QCD, the copula and the marginal distributions of a given pair $(X_{t,j_1}^{i}, X_{t+l,j_2}^{i})$, which gives $d_{QCD}$ the ability to capture any kind of deviation in the dependence structure of the stochastic processes. \\

\section{Fuzzy clustering methods based on the quantile cross-spectral density and PCA}\label{sectionqcdpca}

In this section we introduce two fuzzy clustering procedures based on the distance $d_{QCD}$, namely the QCD-based fuzzy $C$-means clustering model and the QCD-based fuzzy $C$-medoids clustering model, and show how the effectiveness of these models can be significantly improved by applying the PCA transformation over the corresponding QCD-based features and performing clustering in the transformed space. 

\subsection{QCD-based fuzzy $C$-means clustering model (QCD-FCMn)}\label{subsectionqcdfcmn}

As in previous sections, consider a set $S$ of $n$ realizations of multivariate time series $\{\bm X_t^{(1)}, \ldots, \bm X_t^{(n)}\}$ and denote by $\bm \Psi=\{\bm \Psi^{(1)}, \ldots, \bm \Psi^{(n)}\}$ the corresponding vector of estimated quantile cross-spectral densities obtained as indicated in Section~\ref{subsectiondqcd}. Assume that all vectors $\bm \Psi^{(i)}$ have the same length, $2d^2r^2(\lfloor T/2 \rfloor+1)$, being $d$ the number of dimensions, $r$ the number of probability levels, $T$ the series length and $\lfloor \cdot \rfloor$ the floor function. This way, a pairwise dissimilarity matrix can be computed according to \eqref{d.QCD}. In this framework, we propose to perform partitional fuzzy clustering on $S$ by using the QCD-based fuzzy $C$-means clustering model (QCD-FCMn), whose aim is to find a set of centroids $\overline{\bm \Psi}=\{\overline{\bm \Psi}^{(1)}, \ldots, \overline{\bm \Psi}^{(C)}\}$, and the $n \times C$ matrix of fuzzy coefficients, $\bm U=(u_{ic})$, $i=1,\ldots,n$,  $c=1,\ldots,C$, which define the solution of the minimization problem
\begin{equation}\label{qcd_means}
\begin{dcases}
\min_{\overline{\bm \Psi}, \bm U}\sum_{i=1}^{n}\sum_{c=1}^{C}u_{ic}^m 
\norm{\bm \Psi^{(i)}-\overline{\bm \Psi}^{(c)}}^2 \\
 \text{with respect to the constraints:}  \\
\sum_{c=1}^{C}u_{ic}=1, \, u_{ic} \geq 0  \hspace{0.8 cm} \text{(constraints on the membership degree)},  \\
\overline{\bm \Psi}_L \le \overline{\bm \Psi}^{(c)}_k \le \overline{\bm \Psi}_U\hspace{0.8 cm} \text{(possible constraints on $\overline{\bm \Psi}^{(c)}_k$)},
\end{dcases}
\end{equation}
where $u_{ic} \in[0,1]$ represents the membership degree of the $i$-th series in the $c$-th cluster, $\overline{\bm \Psi}^{(c)}$ is the vector of estimated quantile cross-spectral densities with regards to the centroid series for the cluster $c$, $m>1$ is a parameter controlling the fuzziness of the partition, usually referred to as fuzziness parameter, and $\overline{\bm \Psi}^{(c)}_k$ is the $k$-th component of centroid $\overline{\bm \Psi}^{(c)}$. Constraints on $u_{ic}$ are standard requirements in fuzzy clustering. Specifically, that the sum of the membership degrees for each series equals one implies that all of them contribute with the same weight to the clustering process. The fuzziness parameter controls the level of fuzziness considered in the grouping procedure. In the naive case, when $m=1$, we have $u_{ic}=1$ if $c=\underset{{c' \in \{1,\ldots,c\}}}{\operatorname{\argmin}}d^2_{QCD}(\bm \Psi^{(i)}, \overline{\bm \Psi}^{(c')})$ and 0 otherwise so that the crisp version of the procedure is obtained. As the value of $m$ increases, the boundaries between clusters become softer and therefore the grouping is fuzzier. Note that the centroid of a cluster is the mean of all points (in this case, the quantile cross-spectral features describing the MTS), weighted by the degree of belonging to the cluster. Hence, we can think of the centroids as the prototypes of each cluster, i.e, a time series or a feature vector artificially representing the characteristics of the time series belonging to the corresponding cluster with a high membership degree. In \eqref{qcd_means}, $\overline{\bm \Psi}_L$ an $\overline{\bm \Psi}_U$ stand for the possible lower and upper bound of $\overline{\bm \Psi}^{(c)}_k$, respectively. 

The goal of QCD-FCMn is to find a fuzzy partition into $C$ clusters such that the squared QCD-distance between the clusters and their prototypes is minimized. 

The quality of the clustering solution strongly depends on the capability of the distance $d_{QCD}$ to identify different dependence structures. Note that, unlike in a crisp clustering procedure, here the non-stochastic uncertainty inherent to the assignment of series to clusters is incorporated to the procedure by means of the membership degrees. 

By taking into consideration only the membership degree constraints, the constrained optimization problem in \eqref{qcd_means} can be solved by means of the Lagragian multipliers method, given rise to a two-step iterative process. The first step consists of the minimization of the objective function with respect to $u_{ic}$, being $\bm {\overline \Psi}$ fixed.
\begin{equation}\label{updatememmeans}
     u_{ic}=\Bigg[ \sum_{c'=1}^{C} \Bigg ( \frac{\norm{\bm \Psi^{(i)}-\overline{\bm \Psi}^{(c)}}^2}{ \norm{\bm \Psi^{(i)}-\overline{\bm \Psi}^{(c')}}^2}\Bigg )^{\frac{2}{m-1}} \Bigg]^{-1}, \, \, \text{for} \,
    i=1,\ldots,n \, \, \text{and} \, \, c=1,\ldots,C.
\end{equation}

The second step is based on the minimization of the objective function regarding $\bm {\overline \Psi}$, being $u_{ic}$ fixed

\begin{equation}\label{updatecent}
\overline{\bm \Psi}^{(c)}_k=\frac{\sum_{i=1}^{n}u_{ic}^m{\bm \Psi}^{(i)}_k}{\sum_{i=1}^{n}u_{ic}^m}, \, \, \text{for} \,
    k=1,\ldots, d^2r^2(\lfloor T/2 \rfloor+1) \, \, \text{and} \, \, c=1,\ldots,C.
\end{equation}

Note that, in the previous iterative solutions, $\overline{\bm \Psi}^{(c)}_k$ already verifies that $\overline{\bm \Psi}_L \le \overline{\bm \Psi}^{(c)}_k \le \overline{\bm \Psi}_U$, since it inherits the possible constraints of the estimated quantile cross-spectral density features from the observed series, $\bm \Psi^{(i)}$ ($i=1,\ldots,n$), i.e., $\overline{\bm \Psi}_L \le {\bm \Psi}^{(i)}_k \le \overline{\bm \Psi}_U$. Indeed, from the previous inequality we have $\sum_{i=1}^{n}u_{ic}^m\overline{\bm \Psi}_L \le \sum_{i=1}^{n}u_{ic}^m{\bm \Psi}^{(i)}_k \le \sum_{i=1}^{n}u_{ic}^m\overline{\bm \Psi}_U$, which, by dividing by $\sum_{i=1}^{n}u_{ic}^m$, yields $\overline{\bm \Psi}_L \le \overline{\bm \Psi}^{(c)}_k \le \overline{\bm \Psi}_U$.
An outline of the QCD-FCMn clustering algorithm is shown in Algorithm \ref{algorithm1}.

\begin{algorithm}
  \caption{The QCD-based fuzzy $C$-means clustering algorithm (QCD-FCMn)  \label{algorithm1}}
  \begin{algorithmic}[1]
    \State Fix $C$, $m$, \textit{max.iter}, \textit{tol}, \text{a matrix norm} $\norm{\cdot}_M$ 
		\State Set $iter \, =0$ 
		\State Initialize the membership matrix $\bm U = \bm U^{(0)}$
		\Repeat
			\State Set $\bm U_{\text{OLD}}= \bm U$   
			\Comment{Store the current membership matrix}
			\State Compute the centroids $\overline{\bm \Psi}^{(c)}$, $c=1,\ldots,C$, by means of \eqref{updatecent}
			\State Compute $u_{ic}$, $i=1,\ldots,n$, $c=1,\ldots,C$, using (\ref{updatememmeans})
			\Comment{Update the membership matrix}
			\State $iter \, \gets iter \, + 1$
		\Until{ \mbox{ $\norm{\bm U -\bm U_{\text{OLD}}}_M<tol \mbox{ or } iter \, = \, max.iter$ } }
 \end{algorithmic}
\end{algorithm}

\subsection{QCD-based fuzzy $C$-medoids clustering model (QCD-FCMd)}\label{subsectionqcdfcmd}

One natural alternative to QCD-FCMn is the QCD-based fuzzy $C$-medoids clustering model (QCD-FCMd). Following the context of Section \ref{subsectionqcdfcmn}, the goal is now to find the subset of $\bm \Psi$ of size $C$, $\widetilde{\bm \Psi}=\{\widetilde{\bm \Psi}^{(1)}, \ldots, \widetilde{\bm \Psi}^{(C)}\}$, and the $n \times C$ matrix of fuzzy coefficients, $\bm U$, solving the minimization problem
\begin{equation}\label{qcd_med}
\min_{\widetilde{\bm \Psi}, \bm U}\sum_{i=1}^{n}\sum_{c=1}^{C}u_{ic}^m
\norm{\bm \Psi^{(i)}-\widetilde{\bm \Psi}^{(c)}}^2 \, \text{with respect to} \,\sum_{c=1}^{C}u_{ic}=1 \,\text{and} \, u_{ic} \geq 0,
\end{equation}
where $\widetilde{\bm \Psi}^{(c)}$ is the vector of estimated quantile cross-spectral density with regards to the medoid series for the cluster $c$ and the remaining elements are the same as in \eqref{qcd_means}.

By solving the constrained optimization problem in \eqref{qcd_med} with the Lagragian multipliers method, we can obtain an iterative algorithm that alternately optimizes the membership degrees and the medoids. The iterative solution for the membership degrees takes the form \cite{hoppner1999fuzzy}
\begin{equation}\label{updatememmedoids}
u_{ic}=\Bigg[ \sum_{c'=1}^{C} \Bigg ( \frac{\norm{\bm \Psi^{(i)}-\widetilde{\bm \Psi}^{(c)}}^2}{ \norm{\bm \Psi^{(i)}-\widetilde{\bm \Psi}^{(c')}}^2}\Bigg )^{\frac{1}{m-1}} \Bigg]^{-1}, \, \text{for} \,
i=1,\ldots,n \, \text{and} \, c=1,\ldots,C.
\end{equation} 

Once the membership degrees are obtained through \eqref{updatememmedoids}, the $C$ series minimizing \eqref{qcd_med} are selected as new medoids. This two-step procedure is iterated until there is no change in the medoids or a maximum number of iterations is achieved. An outline of the QCD-FCMd clustering algorithm is given in Algorithm \ref{algorithm2}.

\begin{algorithm}
	\caption{The QCD-based fuzzy $C$-medoids clustering algorithm (QCD-FCMd)  \label{algorithm2}}
	\begin{algorithmic}[1]
		\State Fix $C$, $m$ and \textit{max.iter}  
		\State Set $iter \, =0$ 
		\State Pick the initial medoids $\widetilde{\bm \Psi}=\{\widetilde{\bm \Psi}^{(1)}, \ldots, \widetilde{\bm \Psi}^{(C)}\}$
		\Repeat
		\State Set $\widetilde{\bm{\Psi}}_{\text{OLD}}= \widetilde{\bm{\Psi}}$   
		\Comment{Store the current medoids}
		\State Compute $u_{ic}$, $i=1,\ldots,n$, $c=1,\ldots,C$, using (\ref{updatememmedoids})
		\State For each $c \in \{1,\ldots,C\}$, determine the index $j_c \in \{1,\ldots,n\}$ satisfying:
		\Statex \hspace*{0.45cm} $\displaystyle j_c = \argmin_{1 \leq j \leq n} \sum_{i=1}^{n} u_{ic}^m \norm{ \bm{\Psi}^{(i)}-\bm{\Psi}^{(j)} }^2$
		\State \textbf{return} $\widetilde{\bm{\Psi}}^{(c)}= \bm{\Psi}^{(j_c)}$, for $c=1,\ldots,C$  
		\Comment{Update the medoids}
		\State $iter \, \gets iter \, + 1$
		\Until{ \mbox{ $\widetilde{\bm{\Psi}}_{\text{OLD}} = \widetilde{\bm{\Psi}} \mbox{ or } iter \, = \, max.iter$ } }
	\end{algorithmic}
\end{algorithm}

\subsection{Effectiveness of combining the quantile cross-spectral density and PCA}\label{subsectioneffectivenessqcdpca}

This section illustrates the advantages of applying the PCA transformation over the features obtained through QCD in order to perform fuzzy cluster analysis.

Consider two bivariate VAR processes with matrices of coefficients given by $\begin{pmatrix}
0.2 & 0.2 \\
0.2 & 0.2 
\end{pmatrix}$ and $\begin{pmatrix}
-0.2 & -0.2 \\
-0.2 & -0.2 
\end{pmatrix}$ respectively, and a bivariate white noise process. In the three cases, we assume that the error vector follows a standard bivariate normal distribution. We simulated 5 MTS of length $T=500$ from each one the processes, giving rise to a set of 15 MTS. The QCD-based features, that is, the vectors in \eqref{uj1j2}, were extracted from each series in order to perform fuzzy clustering. The set of probability levels $\mathcal{T}=\{0.1, 0.5, 0.9\}$ was considered. This set if often enough for the quantile-based features to give a meaningful description of the underlying process \cite{vilar2018quantile,lafuente2016clustering, lafuente2018robust,oriona2020, oriona2021a, oriona2021b}. Additionally, the PCA transformation was applied over the matrix containing the QCD-based feature vectors as rows. The so-called matrix of scores (giving the position of each observation in the new coordinate system) was obtained and the first $\lfloor 0.12p \rfloor$ principal components were retained, being $p$ the total number of principal components, 15 in this example. The rate 0.12 was chosen due to the fact that it proved very effective in retaining enough information about the original dataset while removing most of the noise for classification tasks (see \cite{oriona2021a}). In fact, we have also tried other values for this parameter, getting worse results in terms of clustering accuracy. The QCD-FCMn algorithm (Algorithm~\ref{algorithm1}) was applied over the original features and over the matrix of scores by setting $C=3$. Four values for the fuzziness parameter $m$ were considered, namely $m=1.5, 1.8, 2, 2.2$. This mechanism was repeated 100 times.

The assessment of both approaches was carried out by means of the fuzzy extension of the Adjusted Rand Index (see Section \ref{subsubsectionfirstassessment}), denoted by FARI. This performance metric is specifically designed to evaluate a fuzzy clustering partition, is bounded between -1 and 1, and the closer to one its value, the better the clustering partition. The partition defined by the generating processes was assumed to be the true partition. Table \ref{tabletoy} contains the average FARI for each of the values of $m$ with both approaches, which are referred to as QCD and QCD-PCA. The results achieved by QCD-PCA were better than the ones attained by using the raw features, specially as the value of $m$ got larger. 

\begin{table}
	\centering
	\begin{tabular}{cccc}
		\hline
		Fuzziness parameter & QCD & QCD-PCA \\ \hline
		$m=1.5$ & 0.747 & \textbf{0.960} \\ 
		$m=1.8$ & 0.511 & \textbf{0.900} \\ 
		$m=2.0$ & 0.402 & \textbf{0.848} \\ 
		$m=2.2$ & 0.322 & \textbf{0.792} \\ 
		\hline
	\end{tabular}
	\caption{Average values of FARI for the fuzzy $C$-means procedures QCD and QCD-PCA according to the fuzziness parameter. For each value of $m$, the best result is shown in bold.}
	\label{tabletoy}
\end{table}

To gain a deeper insight into the behaviour of both approaches, we examined the resulting membership degrees. First, we obtained the average value of the maximum membership degree concerning each of the three clusters (see Table \ref{tabletoy_max_memberships}). The membership degrees generated by QCD-PCA are far larger than those from QCD, thus concluding that QCD-PCA best resolves the level of uncertainty by providing a fuzzy partition with membership values close to one or zero.

\begin{table}
	\centering
	\begin{tabular}{ccccc}
		\hline
		&  &  & QCD & QCD-PCA \\ \hline 
		$m=1.5$ &  & Cluster 1 & 0.915 & \textbf{1} \\ 
		&  & Cluster 2 & 0.916 & \textbf{1} \\ 
		&  & Cluster 3 & 0.916 & \textbf{1} \\ \hline 
		$m=1.8$ &  & Cluster 1 & 0.770 & \textbf{0.996} \\ 
		&  & Cluster 2 & 0.779 & \textbf{0.995} \\ 
		&  & Cluster 3 & 0.771 & \textbf{0.996} \\ \hline 
		$m=2.0$ &  & Cluster 1 & 0.693 & \textbf{0.987} \\ 
		&  & Cluster 2 & 0.687 & \textbf{0.987} \\ 
		&  & Cluster 3 & 0.692 & \textbf{0.991} \\ \hline 
		$m=2.2$ &  & Cluster 1 & 0.620 & \textbf{0.975} \\ 
		&  & Cluster 2 & 0.624 & \textbf{0.978} \\ 
		&  & Cluster 3 & 0.630 & \textbf{0.978} \\ 
		\hline
	\end{tabular}
	\caption{Average maximum membership degrees for fuzzy $C$-means procedures QCD and QCD-PCA according to the fuzziness parameter. For each value of $m$, the best result is shown in bold.}
	\label{tabletoy_max_memberships}
\end{table}

Then, we recorded the proportion of times that the fuzzy partition produced by both approaches resulted in the true crisp partition. To this end, we decided to assign the $i$-th MTS to the $c$-th cluster according to the two criteria: $u_{ic}>0.8$ and $u_{ic}>0.6$, thus evaluating both approaches under different degrees of requirement in the assignment rule. The corresponding rates of correct identification are shown in Table \ref{tabletoyrates}. We can see that, in both cases, QCD-PCA outperformed QCD by a large extent. The latter was unable to get the true partition when the value of $m$ was 2 or 2.2. These results suggest that the solutions attained by QCD-PCA are far closer to the true partition. Note that it is reasonable that both procedures decreased their performance when increasing the value of $m$, since higher values of $m$ imply softer partitions. 

\begin{table}
\centering
	\begin{tabular}{cccccc} \hline 
		& \multicolumn{2}{c}{$u_{ic}>0.8$} & \multicolumn{2}{c}{$u_{ic}>0.6$}  \\ \hline 
		$m$ & QCD     & QCD-PCA    & QCD     & QCD-PCA      \\ \hline 
		$m=1.5$	&    0.17     &    \textbf{0.92}        &    0.87     &     \textbf{0.92}       &      \\
		$m=1.8$ &   0   &    \textbf{0.54}        &    0.47     &      \textbf{0.85}      &       \\
		$m=2.0$ & 0  &      \textbf{0.30}      &    0.04     &     \textbf{0.82}       &          \\
		$m=2.2$	& 0    &       \textbf{0.14}     &   0      &     \textbf{0.73}       &      \\ \hline    
	\end{tabular}
	\caption{Rates of correctly identified crisp partitions for fuzzy $C$-means procedures QCD and QCD-PCA according to the fuzziness parameter and two different cutoff values. For each value of $m$ and the cutoff, the best result is shown in bold.}
	\label{tabletoyrates}
\end{table}

Thus, one can conclude that, generally, when PCA transformation is not considered, the resulting partitions show a high degree of overlap between clusters, thus giving little informative solutions. On the other hand, by applying PCA, large membership degrees are usually attained, which provides more reliable solutions. This holds true for the most common values of $m$ used in practice. 

It is worth highlighting that the results achieved by QCD-PCA in this toy example are likely to be attributable to the noise reduction and higher stability resulting from the PCA transformation. PCA is often used as a preprocessing step to unsupervised classification \cite{everitt2011cluster}, usually providing a robustification to the clustering technique \cite{ben2003detecting}. Indeed, several works show its usefulness in different application domains \cite{raychaudhuri1999principal, xue2011using, ding2004k, gaitani2010using}. Note that, due to the definition of the smoothed CCR-periodogram in \eqref{smoothedCCR}, the QCD-based features are highly correlated. Therefore, by considering the raw features, some variables could get a higher weight than others in the distance computation, thus creating a bias in the clustering algorithm. The PCA transformation avoids this problem by removing the underlying correlation between features, thus making the grouping process easier.

The better performance of QCD-PCA over QCD holds generally true whatever the generating processes. Specifically, in the simulated scenarios considered in Section \ref{sectionsimulationstudy}, QCD-PCA achieved substantially better results than QCD. For this reason, from now on, the distance $d_{QCD}$ and the  clustering procedures QCD-FCMn and QCD-FCMd are going to refer to the PCA-transformed features rather than the original features, although we maintain the notation for the sake of simplicity and readability. More precisely, given a set of $n$ MTS, $\{\bm X_t^{(1)}, \ldots, \bm X_t^{(n)}\}$, the original QCD-based features are extracted from each series, thus providing the set $\bm \Psi=\{\bm \Psi^{(1)}, \ldots, \bm \Psi^{(n)}\}$. These vectors are transformed by means of PCA, obtaining the set of score vectors $\bm \Psi_{PCA}=\{\bm \Psi_{PCA}^{(1)}, \ldots, \bm \Psi_{PCA}^{(n)}\}$. This is the set subject to clustering by means of Algorithms \ref{algorithm1} and \ref{algorithm2}, and the corresponding distance is $d_{QCD_{PCA}}(\bm X_t^{(1)}, \bm X_t^{(2)})=\norm{\bm \Psi_{PCA}^{(1)}-\bm \Psi_{PCA}^{(2)}}^2$. The subscript $PCA$ is removed from now on.  

\section{Simulation study}\label{sectionsimulationstudy}

In this section, we carry out a set of simulations with the aim of evaluating the performance of $d_{QCD}$ in different scenarios of fuzzy clustering of MTS. Firstly we describe some alternative metrics that we have  considered for comparison purposes. Then we explain the two ways in which the assessment task was performed, along with the corresponding simulation mechanism and results. 

\subsection{Alternative metrics}\label{subsectioncompetitors}

To shed light on the performance of $d_{QCD}$ in a fuzzy clustering context, it was compared with some other clustering models based on alternative dissimilarities. Note that, according to the fuzzy approach based on features extracted from an MTS, a fuzzy $C$-medoids model and a fuzzy $C$-means model can be formalized as the minimization problems in \eqref{qcd_med} and \eqref{qcd_means}, respectively, only by replacing $\widetilde{\bm \Psi}$, $\widetilde{\bm \Psi}^{(c)}$, $\bm \Psi^{(i)}$, $\overline{\bm \Psi}$, $\overline{\bm \Psi}^{(c)}$, $\overline{\bm \Psi}_L$,  $\overline{\bm \Psi}_U$ by $\widetilde{\bm \varphi}$, $\widetilde{\bm \varphi}^{(c)}$, $\bm \varphi^{(i)}$, $\overline{\bm \varphi}$, $\overline{\bm \varphi}^{(c)}$, $\overline{\bm \varphi}_L$,  $\overline{\bm \varphi}_U$, where $\bm \varphi^{(i)}$ represents the vector of estimated features for the $i$-th series, $i = 1\ldots, n$, and the remaining terms are defined analogously. In the same way, the iterative solutions are obtained through \eqref{updatememmedoids}, \eqref{updatememmeans} and \eqref{updatecent} by considering the corresponding features. 

The alternative approaches are described below:
\begin{itemize}
 \item \textit{Wavelet-based features}. \cite{d2012wavelets} introduced a squared Euclidean distance between wavelet features of two MTS, specifically between estimates of wavelet variances and wavelet correlations. The estimates are obtained through the maximum overlap discrete wavelet transform, which requires choosing a wavelet filter of a given length and a number of scales. Thus, in this case the vector $\bm \varphi^{(i)}$ contains estimates of wavelet variances and wavelet correlations of a given MTS. The corresponding methods are referred to as Wavelet-based fuzzy $C$-medoids clustering model (W-FCMd) and Wavelet-based fuzzy $C$-means clustering model (W-FCMn). After performing some brief preliminary analyses, we reached the conclusion that the wavelet filter of length 4 of the Daubechies family, DB4, along with the maximum allowable number of scales, were the choices that led to the best average results in the considered simulation scenarios (see Sections \ref{subsubsectionfirstassessment} and \ref{subsubsectionsecondassessment}). Hence, they were the hyperparameters chosen for the simulation study.
 \item \textit{Correlation-based features}. In the univariate framework, \cite{d2009autocorrelation} proposed a fuzzy procedure in which the extracted features are the estimated autocorrelations of a UTS for lags between 1 and fixed $l$. Here we propose to generalize this approach to a multivariate context. This generalization is straightforward. Given an MTS, we fix a lag $l$ and compute estimates of the autocorrelations up to lag $l$ for each component (UTS) and of the cross-correlations up to lag $l$ between each pair of different components. This set of features defines the vector $\bm {\varphi}^{(i)}$ describing the $i$-th MTS and is used to perform clustering. We call the corresponding approaches Correlation-based fuzzy $C$-medoids clustering model (C-FCMd) and Correlation-based fuzzy $C$-means clustering model (C-FCMn). The hyperparameter $l$ was set to $l=1$ throughout the simulation study, as the majority of the considered generating processes contain only one significant lag (see Sections \ref{subsubsectionfirstassessment} and \ref{subsubsectionsecondassessment}). 
 \item \textit{A versatile approach based on features of different nature}. \cite{wang2007structure} provided a two-step procedure for MTS clustering. First, a set of features of different nature (skewness, kurtosis, nonlinear structure...) are extracted from each UTS within the MTS and stored in a vector. Second, these vectors are concatenated with each other in order to construct a unique vector describing the $i$-th MTS, $\bm \varphi^{(i)}$. We refer to the corresponding approaches as Features-based fuzzy $C$-medoids clustering model (F-FCMd) and Features-based fuzzy $C$-means clustering model (F-FCMn).
 \item \textit{VPCA metric}. It was proposed by \cite{he2018unsupervised}. The procedure is based on PCA and a spatial weighted distance matrix. The grouping is carried out over a set of matrices which are constructed applying dimensionality reduction techniques to the original set of MTS. Using a distance between matrices called SWMD, a fuzzy $C$-means approach is proposed (see Section 3.B in \cite{he2018unsupervised}). Its extension to the case of a fuzzy-C medoids is straightforward. Note that this method does not pertain to the fuzzy framework based on extracted features, so here there is no vector $\bm \varphi^{(i)}$. The approaches are called the VPCA-based fuzzy $C$-medoids clustering model (VPCA-FCMd) and the VPCA-based fuzzy $C$-means clustering model (VPCA-FCMn).

\end{itemize}

It is worth remarking that we did not consider the well-known DTW distance in an alternative procedure. This was due to the fact that, in our previous work \cite{oriona2020}, this distance proved totally useless for grouping MTS according to the generating processes in a crisp clustering framework. Indeed, we have made some proofs with this metric in a fuzzy context and it attained very poor results.

\subsection{Experimental design and results}\label{subsectionexperimentaldesign}

A broad simulation study was carried out to evaluate the performance of the proposed methods, QCD-FCMn and QCD-FCMd. We intended to make the evaluation process so that general conclusions about the performance of both approaches can be reached. To this end, we designed two different assessment schemes. The first one contains scenarios with three different groups of MTS, and is aimed at evaluating the capability of the procedures in assigning high (low) membership if a given MTS pertains (not pertains) to a given cluster. The second one consists of scenarios with two different groups of MTS and an MTS which does not belong to any of the groups. It assesses again the membership degrees of the series in the two groups but also that the switching series is not assigned to any of the clusters with high membership. The latter scheme utilises a cutoff value determining whether or not a membership degree in a given group is enough to assign the corresponding MTS to that cluster.

\subsubsection{First assessment scheme}\label{subsubsectionfirstassessment}

In this section, the performance of QCD-FCMn and QCD-FCMd is analysed by means of three different simulated scenarios. We attempted to recreate scenarios with a wide variety of generating processes (linear, nonlinear and conditional heteroskedastic), number of dimensions, and series lengths. Each scenario consisted of three clusters (i.e., generating models) with five MTS each, defining the true clustering partition. The generating models concerning each class of processes are given below.

\noindent \textbf{Scenario 1}. Fuzzy clustering of VARMA processes. \\

\noindent (a) VAR(1)

\begin{equation*}
\begin{pmatrix}
X_{t,1}  \\
X_{t,2} \\ 
X_{t, 3}
\end{pmatrix} = 
\begin{pmatrix}
0.4 & -0.3 & 0.9  \\
0.4 & 0.3 & -0.1 \\ 
0.5 & 0.3 & -0.2
\end{pmatrix} 
\begin{pmatrix}
X_{t-1,1}   \\
X_{t-1,2}\\ 
X_{t-1,3} 
\end{pmatrix} +
\begin{pmatrix}
\epsilon_{t,1}   \\
\epsilon_{t,2}\\ 
\epsilon_{t,3} 
\end{pmatrix},
\end{equation*}

\noindent (b) VMA(1)
\begin{equation*}
\begin{pmatrix}
X_{t,1}  \\
X_{t,2} \\ 
X_{t, 3}
\end{pmatrix} = 
\begin{pmatrix}
0.3 & -0.7 & -0.9  \\
0.2 & 0.3 & 0.1 \\ 
0.2 & 0.1 & -0.3
\end{pmatrix} 
\begin{pmatrix}
X_{t-1,1}   \\
X_{t-1,2}\\ 
X_{t-1,3} 
\end{pmatrix} +
\begin{pmatrix}
\epsilon_{t,1}   \\
\epsilon_{t,2}\\ 
\epsilon_{t,3} 
\end{pmatrix},
\end{equation*}

\noindent (c) VARMA(1,1)

\begin{eqnarray*}
	\begin{pmatrix}
		X_{t,1}  \\
		X_{t,2} \\ 
		X_{t, 3}
	\end{pmatrix} & = &   
	\begin{pmatrix}
		0.6 & 0.5 & 0  \\
		-0.4 & 0.5 & 0.3 \\ 
		0 & -0.5 & 0.7
	\end{pmatrix} 
	\begin{pmatrix}
		X_{t-1,1}   \\
		X_{t-1,2}\\ 
		X_{t-1,3} 
	\end{pmatrix} + \\
	& & 
	\begin{pmatrix}
		0.6 & 0.5 & 0  \\
		-0.4 & 0.5 & 0.3 \\ 
		0 & -0.5 & 0.7
	\end{pmatrix} 
	\begin{pmatrix}
		\epsilon_{t-1,1}   \\
		\epsilon_{t-1,2}\\ 
		\epsilon_{t-1,3} 
	\end{pmatrix} +  
	\begin{pmatrix}
		\epsilon_{t,1}   \\
		\epsilon_{t,2}\\ 
		\epsilon_{t,3} 
	\end{pmatrix}.
\end{eqnarray*}

\noindent \textbf{Scenario 2}. Fuzzy clustering of nonlinear processes. \\

\noindent (a) NVAR (nonlinear vector autoregressive process)
\begin{equation*}
\begin{pmatrix}
X_{t,1}  \\
X_{t,2} 
\end{pmatrix} = 
\begin{pmatrix}
0.7|X_{t-1,1}|/(|X_{t-1,2}|+1) \\ 
0.7|X_{t-1,2}|/(|X_{t-1,1}|+1)
\end{pmatrix} 
+
\begin{pmatrix}
\epsilon_{t,1}   \\
\epsilon_{t,2}
\end{pmatrix},
\end{equation*}

\noindent (b) TAR (threshold autoregressive process)
\begin{equation*}
\begin{pmatrix}
X_{t,1}  \\
X_{t,2} 
\end{pmatrix} = 
\begin{pmatrix}
0.9X_{t-1,2}I_{\{|X_{t-1,1}| \leq 1\}} -0.3X_{t-1,1}I_{\{|X_{t-1,1}| > 1\}} \\
0.9X_{t-1,1}I_{\{|X_{t-1,2}| \leq 1\}} -0.3X_{t-1,2}I_{\{|X_{t-1,2}| > 1\}} 
\end{pmatrix} 
+
\begin{pmatrix}
\epsilon_{t,1}   \\
\epsilon_{t,2}
\end{pmatrix},
\end{equation*}

\noindent (c) BL (bilinear process)
\begin{equation*}
\begin{pmatrix}
X_{t,1}  \\
X_{t,2} 
\end{pmatrix} = 
\begin{pmatrix}
0.7X_{t-1,1}\epsilon_{t-2,2}\\
0.7X_{t-1,2}\epsilon_{t-2,1}
\end{pmatrix} 
+
\begin{pmatrix}
\epsilon_{t,1}   \\
\epsilon_{t,2}
\end{pmatrix}.
\end{equation*}

\noindent \textbf{Scenario 3}. Fuzzy clustering of dynamic conditional correlation processes. Consider $(X_{t,1}, X_{t,2})^\intercal=(a_{t,1}, a_{t, 2})^\intercal=(\sigma_{t,1}\epsilon_{t,1}, \sigma_{t, 2}\epsilon_{t,2})^\intercal$, denoting $\intercal$ the transpose vector. The data generating process consists of two GARCH models. Specifically, 
\begin{equation*}
\sigma_{t,1}^2=0.01+0.05a_{t-1,1}^2+0.94\sigma^2_{t-1,1},
\end{equation*}
\begin{equation*}
\sigma_{t,2}^2=0.5+0.2a_{t-1,2}^2+0.5\sigma^2_{t-1,2},
\end{equation*}
\begin{equation*}
\begin{pmatrix}
\epsilon_{t,1} \\
\epsilon_{t,2} 
\end{pmatrix} 
\sim N
\left [ \begin{pmatrix}
0 \\
0
\end{pmatrix},\begin{pmatrix}
1 & \rho_t \\
\rho_t & 1
\end{pmatrix} \right].
\end{equation*}

The correlation between the standardized shocks, $\rho_t$, is given by the following expressions:

\noindent (a) Piecewise constant correlation
\begin{equation*}
\rho_t=0.9I_{\{t\leq(T/2)\}}-0.7I_{\{t>(T/2)\}}
\end{equation*} 

\noindent (b) Constant correlation
\begin{equation*}
\rho_t=0.5,
\end{equation*}

\noindent (c) Piecewise constant correlation
\begin{equation*}
\rho_t=0.9I_{\{t\leq(T/2)\}}-0.2I_{\{t>(T/2)\}}.
\end{equation*}

The error vector in Scenarios 1 and 2 follows a multivariate standard Gaussian distribution. 

VARMA models are broadly used in many fields but the determination of the models order is quite complex since fixing orders too small leads to inconsistent estimators whereas too large orders produce less accurate predictions. Note that our approach does not require prior modeling. Scenario~2 consists of multivariate extensions of univariate NAR, TAR, and BL processes proposed in \cite{Zhang2001}. Nonlinear UTS arise in several application fields \cite{granger1993modelling, granger1978introduction, tong2009threshold}. Although nonlinear MTS have received much less attention than linear ones, there exist some fields as neurophysiology \cite{pereda2005nonlinear} and economy \cite{Koop1996} in which nonlinear analysis of MTS has proven to be critical. Thus, a good fuzzy clustering method should be able to specify proper membership degrees between different nonlinear generating processes. Scenario~3 is based on Scenario 2 in \cite{oriona2020}, which is in turn motivated by a simulation study in the landmark work \cite{engle2002dynamic}, where dynamic conditional correlation models are introduced. Multivariate GARCH models have been comprehensively investigated over the last decades (an extensive survey is offered in \cite{bauwens2006multivariate}). Specifically, estimation of dynamic conditional correlation models has been widely applied to financial series of different nature \cite{ku2007application, naoui2010dynamic, kuper2007dynamic}. Furthermore, we have decided to include in Scenario~3 both positive and negative correlations, since it has been shown that some financial quantities are either positive or negative correlated depending on the period \cite{andersson2008does}. Some of the generating processes in Scenarios 1, 2 and 3 have already been used either for clustering \cite{oriona2020} or classification \cite{oriona2021a} purposes.

We considered different values for the series length, namely $T\in \{100, 150, 200\}$ in Scenario 1, $T\in \{300, 400, 500\}$ in Scenario 2 and $T\in \{1000, 1500, 2000\}$ in Scenario 3 in order to study its effect in the proposed approaches. Note that, as the three scenarios contain very distinct types of processes, it is logical that very different values of $T$ are needed in order to make an appropriate evaluation. Particularly, we considered specially large values of $T$ in Scenario 3. However, this is not necessarily a drawback, as these sample sizes are offered encountered in real MTS fitted by means of dynamic conditional correlation models \cite{ku2007application, kuper2007dynamic}. Indeed multivariate series of stock returns and other related financial quantities, which consist of measures of daily or even intra-daily data, are one common example of series fitted through this class  of models.

The fuzziness parameter $m$ plays a crucial role in the obtained clustering solution. When $m=1$, the crisp version of either fuzzy $C$-means or fuzzy $C$-medoids is obtained. On the other hand, excessively large values of $m$ result in a partition with all memberships close to $1/C$, thus having a large degree of overlap between clusters. As a consequence, choosing these values for $m$ is not recommended \cite{arabie1981overlapping}. There is a broad range of literature on determining proper values for the fuzziness parameter appropriately. For instance, \cite{bezdek2013pattern} showed that values of $m$ between $1.5$ and $2.5$ are typically a good choice for the fuzzy-C means clustering algorithm. This is also confirmed by \cite{cannon1986efficient, hall1992comparison}. However, there seems to be no consensus about the optimal value for $m$ (see discussion in Section 3.1.6 of \cite{maharaj2011fuzzy}). In the context of time series clustering, the majority of works consider values of $m$ between 1.3 and 2.6 when performing simulation studies \cite{vilar2018quantile, d2012wavelets,d2009autocorrelation, maharaj2011fuzzy, lafuente2018robust}. Based on the previous considerations, we have decided to take into consideration the values $m=1.5$, $1.8$, $2$ and $2.2$.

Given a scenario, a value for $m$ and a value for $T$, 200  simulations were performed. In each trial, we applied the QCD-FCMn method, as well as the fuzzy $C$-means versions of the competitors in Section \ref{subsectioncompetitors}. The same procedure was repeated regarding the QCD-FCMd technique and the corresponding fuzzy-$C$-medoids variants. The set of probability levels $\mathcal{T}=\{0.1, 0.5, 0.9\}$ and the first $\lfloor 0.12p \rfloor$ principal components, being $p$ the total number of principal components, were used to compute the QCD-based features. The number of clusters was set to $C=3$. The effectiveness of each clustering procedure was measured by means of the fuzzy extension of the Adjusted Rand Index (FARI) devised by \cite{campello2007fuzzy}. This index compares the true partition with the membership matrix resulting from a clustering algorithm. It is based on the original definition of the Adjusted Rand Index and some notions of the fuzzy set theory. The resulting index is also bounded between -1 and 1, as the original Adjusted Rand Index, indicating values close to 1 a more accurate clustering solution. One of the biggest advantages of the FARI is that it allows to elude the loss of information which is unavoidable when using the original formulation of the index to assess the quality of a fuzzy partition; e.g, by transforming the membership matrix into a crisp partition. This way, two membership matrices giving rise to the same crisp partition can be compared with one another, thus making the comparison process fairer. Computation of FARI requires the selection of a triangular norm (see Section 3.2 in \cite{campello2007fuzzy}) and the corresponding triangular conorm (the dual norm of the former). As the triangular norm, we have considered the minimum norm, which is a common choice in fuzzy logic \cite{bezdek2013pattern}. The respective triangular conorm is the maximum norm.

\subsubsection*{Results}

Table \ref{fuzzycmeanstable} shows the average values of the FARI for the fuzzy $C$-means procedures. We can see that all the methods decreased their performance as the value of the parameter $m$ increased. This is reasonable and expected since we are considering scenarios with three well-defined clusters. As it was already mentioned, small values of $m$ lead to near-crisp partitions. On the contrary, larger values of $m$ imply smoothing the boundary between clusters and make the classification fuzzier, thus decreasing the value of the FARI. 

\begin{table}
	\centering
	\begin{tabular}{ccccccc}
		\hline
		& & QCD-FCMn & W-FCMn & C-FCMn & F-FCMn & VPCA-FCMn \\ \hline 
		\multicolumn{7}{l}{Scenario 1} \\  \hline
		$T=100$ & $m=1.5$ & 0.952 & 0.536 & \textbf{0.975} & 0.309 & 0.027 \\ 
		& $m=1.8$ & \textbf{0.910} & 0.426 & 0.886 & 0.296 & 0.002 \\ 
		& $m=2.0$ & \textbf{0.867} & 0.364 & 0.810 & 0.286 & 0 \\ 
		& $m=2.2$ & \textbf{0.818} & 0.312 & 0.734 & 0.274 & 0 \\  \cline{2-7}
		$T=150$ & $m=1.5$ & \textbf{0.991} & 0.865 & 0.990& 0.369 & 0.026 \\ 
		& $m=1.8$ & \textbf{0.961} & 0.728 & 0.932 & 0.354 & 0.001 \\ 
		& $m=2.0$ & \textbf{0.927} & 0.641 & 0.872 & 0.342 & 0 \\ 
		& $m=2.2$ & \textbf{0.886} & 0.564 & 0.806 & 0.327 & 0 \\   \cline{2-7}
		$T=200$ & $m=1.5$ & \textbf{0.995} & 0.931 & 0.994 & 0.376 & 0.026 \\ 
		& $m=1.8$ & \textbf{0.973} & 0.813 & 0.950 & 0.365 & 0.001 \\ 
		& $m=2.0$ & \textbf{0.946} & 0.729 & 0.898 & 0.355 & 0 \\ 
		& $m=2.2$ & \textbf{0.910} & 0.652 & 0.839 & 0.342 & 0 \\ \hline 
		\multicolumn{7}{l}{Scenario 2} \\  \hline
		$T=300$ & $m=1.5$ & \textbf{0.852} & 0.703 & 0.694 & 0.398 & 0.043 \\ 
		& $m=1.8$ & \textbf{0.811} & 0.626 & 0.638 & 0.380 & 0.002 \\ 
		& $m=2.0$ & \textbf{0.772} & 0.573 & 0.594 & 0.364 & 0 \\ 
		& $m=2.2$ & \textbf{0.733} & 0.521 & 0.549 & 0.348 & 0 \\  \cline{2-7}
		$T=400$ & $m=1.5$ & \textbf{0.931} & 0.839 & 0.764 & 0.405 & 0.046 \\ 
		& $m=1.8$ & \textbf{0.892} & 0.744 & 0.700 & 0.388 & 0.002 \\ 
		& $m=2.0$ & \textbf{0.855} & 0.681 & 0.654 & 0.374 & 0 \\ 
		& $m=2.2$ & \textbf{0.812} & 0.622 & 0.607 & 0.359 & 0 \\ \cline{2-7}
		$T=500$ & $m=1.5$ & \textbf{0.962} & 0.876 & 0.790 & 0.407 & 0.058 \\ 
		& $m=1.8$ & \textbf{0.929} & 0.784 & 0.726 & 0.393 & 0.002 \\ 
		& $m=2.0$ & \textbf{0.896} & 0.723 & 0.681 & 0.380 & 0 \\ 
		& $m=2.2$ & \textbf{0.856} & 0.664 & 0.635 & 0.365 & 0 \\ \hline 
		\multicolumn{7}{l}{Scenario 3} \\  \hline
		$T=1000$ & $m=1.5$ & \textbf{0.631} & 0.402 & 0.568 & 0.003 & 0.002 \\ 
		& $m=1.8$ & \textbf{0.569} & 0.340 & 0.499 & 0.002 & 0 \\ 
		& $m=2.0$ & \textbf{0.522} & 0.298 & 0.451 & 0.002 & 0 \\ 
		& $m=2.2$ & \textbf{0.475} & 0.261 & 0.405 & 0.001 & 0 \\ \cline{2-7}
		$T=1500$ & $m=1.5$ & \textbf{0.783} & 0.482 & 0.622 & -0.010 & 0.002 \\ 
		& $m=1.8$ & \textbf{0.701} & 0.414 & 0.556 & -0.009 & 0 \\ 
		& $m=2.0$ & \textbf{0.643} & 0.369 & 0.509 & -0.008 & 0 \\ 
		& $m=2.2$ & \textbf{0.590} & 0.326 & 0.463 & -0.007 & 0 \\  \cline{2-7}
		$T=2000$ & $m=1.5$ & \textbf{0.863} & 0.575 & 0.718 & 0.009 & 0 \\ 
		& $m=1.8$ & \textbf{0.782} & 0.493 & 0.643 & 0.006 & 0 \\ 
		& $m=2.0$ & \textbf{0.721} & 0.438 & 0.589 & 0.005 & 0 \\ 
		& $m=2.2$ & \textbf{0.660} & 0.391 & 0.535 & 0.005 & 0 \\ 
		\hline
	\end{tabular}
	\caption{Average FARI obtained by the fuzzy $C$-means clustering procedures.}
	\label{fuzzycmeanstable}
\end{table}

The algorithm QCD-FCMn achieved the best scores in all the considered setups except for Scenario 1 with $T=100$ and $m=1.5$, where it was slightly outperformed by the correlation-based approach C-FCMn. In Scenario 1, these two methods obtained very similar scores for all values of $T$ and $m=1.5$, $1.8$. However, for $m=2$, $2.2$, the difference in favour of QCD-FCMn got significant. The wavelet-based technique also obtained acceptable scores in this scenario, particularly for the largest value of $T$ and the smallest values of $m$.

As for Scenario 2, the wavelet and the correlation-based approaches attained similar results, the former outperforming the latter to a small extent as the series length increased. Both techniques were clearly defeated by QCD-FMn, which attained, in all the setups, an average FARI at least 0.08 points better than either. With regards to Scenario 3, the proposed method also beat W-FCMn and C-FCMn by a considerable degree. The difference with respect to the correlation-based procedure, the second best performing method, was smaller for $T=1000$, but substantial for $T=1500$ and $2000$. 

The remaining methods F-FCMn and VPCA-FCMn attained in general poor results. The former achieved an average FARI above 0.30 in almost all the settings of Scenarios 1 and 2, thus indicating that it was able to distinguish between generating processes to some  extent. However, it did not show a significant improvement when larger series were considered. This is probably due to the fact that some of the features taken into consideration by this technique are useless to differentiate between underlying dependence structures, thus making the clustering process noisy. Indeed, in Scenario 3, F-FCMn did not perform better than choosing a membership matrix at random. The procedure based on the spatial weighted distance matrix VPCA-FCMn, got by far the worst results between the five considered algorithms. It was unable to reach meaningful conclusions in all the situations, always obtaining an average FARI near to zero. It is clear from the results that this approach is not appropriate to perform fuzzy clustering based on generating processes. For this reason, we decided to not take this method into account for further analysis.

It is worth remarking that, according to the results in Table \ref{fuzzycmeanstable}, QCD-FCMn was the less affected approach by increasing the value of $m$. Note that, whereas some methods as the wavelet-based procedure usually decreased the average FARI by far more than 0.10 when a given value of $m$ is replaced by the next, the maximum variation of QCD-FCMn was 0.082 in Scenario 3 with $T=1500$, when moving from $m=1.5$ to $m=1.8$ (in the remaining settings, the variation was far less than that). The stability exhibited by QCD-FCMn against the modification of the values of $m$ is a very beneficial property of the proposed approach. As already stated, the fuzziness parameter plays a pivotal role in the quality of the clustering solution, and has to be set in advance in practical applications, usually without guarantees about the rightness of the choice. The devised algorithm QCD-FCMn gets around this limitation, ensuring a high probability of meaningful results whatever the value of $m$. Thus, it could be seen as a sure bet to be used in real clustering problems.

\begin{figure}
	\centering
	\includegraphics[width=1\textwidth]{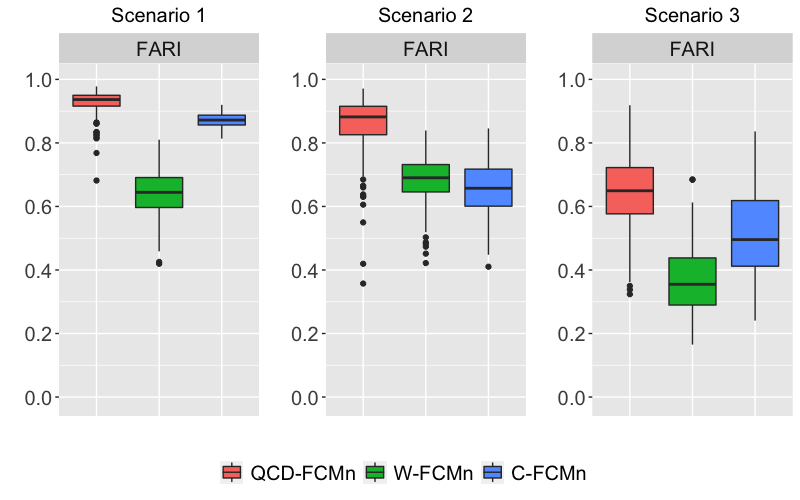}
	\caption{Boxplots of FARI index according to the 200 trials of the simulation procedure. The considered length was $T=150$ for Scenario 1, $T=400$ for Scenario 2 and $T=1500$ for Scenario 3. The fuzziness parameter was set to $m=2$.}
	\label{boxplots}
\end{figure}

In order to gain illustrative insights into the previous results, Figure \ref{boxplots} displays the boxplots based on the FARI according to the 200 simulation trials for intermediate values of $T$ (150, 400 and 1500 for Scenarios 1, 2 and 3, respectively) and $m=2$. We depicted the results only for the three best performing approaches according to Table \ref{fuzzycmeanstable}, QCD-FCMn, W-FCMn and C-FCMn. The superiority of QCD-FCMn over the remaining strategies is obvious from Figure \ref{boxplots}. Additionally, the plots give us an idea about the variability of the results associated with each procedure. In Scenario 1, the results of QCD-FCMn and C-FCMn showed very little dispersion in comparison with those of W-FCMn. On the contrary, in Scenarios 2 and 3, the approaches exhibiting less variability were QCD-FCMn and W-FCMn, whereas the correlation-based technique C-FCMn displayed the most. A similar situation arises when considering the rest of the values for $T$ and $m$. Hence, the proposed algorithm also has the desirable property of giving less variable results than the considered competitors. 

The results involving the fuzzy $C$-medoids versions of the approaches are given in Table \ref{fuzzycmedoidstable}. Generally, the fuzzy $C$-medoids algorithms showed worse effectiveness than the fuzzy $C$-means algorithms. Aside from this fact, some interesting conclusions can be reached from Table \ref{fuzzycmedoidstable}. On the one hand, all the approaches substantially decreased their performance in Scenario 3 as compared with their fuzzy $C$-means counterparts. A different story went for Scenario 1 and 2. In these scenarios, whereas W-FCMn and C-FCMn significantly worsened their scores in comparison with those of Table \ref{fuzzycmeanstable}, QCD-FCMn barely suffered. Indeed, whereas the former approaches decreased in some setups their average FARI by more than 0.15 when changing the algorithm, the greatest decrease for QCD-FCMn, occuring in Scenario 2 when $T=500$ and $m=2.2$, was 0.043. Consequently, the proposed procedure is the one showing the most robustness against a change in the considered clustering algorithm. 

\begin{table}
\centering
\begin{tabular}{cccccc}
  \hline
  &  & QCD-FCMd & W-FCMd & C-FCMd & F-FCMd \\  \hline
 \multicolumn{6}{l}{Scenario 1}  \\ \hline
 $T=100$ & $m=1.5$ & \textbf{0.957} & 0.453 & 0.861 & 0.327 \\ 
 & $m=1.8$ & \textbf{0.900} & 0.342 & 0.706 & 0.301 \\ 
 & $m=2.0$ & \textbf{0.852}& 0.289 & 0.645 & 0.279 \\ 
 & $m=2.2$ & \textbf{0.801} & 0.251 & 0.585 & 0.262 \\ \cline{2-6}
 $T=150$ & $m=1.5$ & \textbf{0.988} & 0.774 & 0.910 & 0.386 \\ 
 & $m=1.8$ & \textbf{0.952} & 0.609 & 0.761 & 0.356 \\ 
 & $m=2.0$ & \textbf{0.912} & 0.522 & 0.723 & 0.335 \\ 
 & $m=2.2$ & \textbf{0.872} & 0.446 & 0.684 & 0.317 \\ \cline{2-6}
 $T=200$ & $m=1.5$ & \textbf{0.991} & 0.862 & 0.890 & 0.394 \\ 
 & $m=1.8$ & \textbf{0.951} & 0.717 & 0.779 & 0.373 \\ 
 & $m=2.0$ & \textbf{0.927} & 0.628 & 0.726 & 0.350 \\ 
 & $m=2.2$ & \textbf{0.894} & 0.543 & 0.706 & 0.330 \\ \hline
  \multicolumn{6}{l}{Scenario 2}  \\ \hline
 $T=300$ & $m=1.5$ & \textbf{0.837} & 0.624 & 0.560 & 0.437 \\ 
 & $m=1.8$ & \textbf{0.799} & 0.524 & 0.518 & 0.386 \\ 
 & $m=2.0$ & \textbf{0.747} & 0.461 & 0.507 & 0.356 \\ 
 & $m=2.2$ & \textbf{0.705} & 0.409 & 0.466 & 0.330 \\  \cline{2-6}
 $T=400$ & $m=1.5$ & \textbf{0.913} & 0.692 & 0.593 & 0.491 \\ 
 & $m=1.8$ & \textbf{0.863} & 0.574 & 0.559 & 0.409 \\ 
 & $m=2.0$ & \textbf{0.818} & 0.528 & 0.540 & 0.378 \\ 
 & $m=2.2$ & \textbf{0.781} & 0.458 & 0.496 & 0.351 \\  \cline{2-6}
 $T=500$ & $m=1.5$ & \textbf{0.936} & 0.671 & 0.625 & 0.558 \\ 
 & $m=1.8$ & \textbf{0.917} & 0.594 & 0.601 & 0.473 \\ 
 & $m=2.0$ & \textbf{0.868} & 0.504 & 0.571 & 0.421 \\ 
 & $m=2.2$ & \textbf{0.813} & 0.477 & 0.517 & 0.380 \\  \hline
   \multicolumn{6}{l}{Scenario 3}  \\ \hline
 $T=1000$ & $m=1.5$ & \textbf{0.537} & 0.341 & 0.520 & -0.002 \\ 
 & $m=1.8$ & \textbf{0.448} & 0.246 & 0.411 & 0.002 \\ 
 & $m=2.0$ & \textbf{0.382} & 0.204 & 0.328 & 0.003 \\ 
 & $m=2.2$ & \textbf{0.353} & 0.170 & 0.271 & 0.003 \\  \cline{2-6}
 $T=1500$ & $m=1.5$ & \textbf{0.656} & 0.408 & 0.591 & -0.010 \\ 
 & $m=1.8$ & \textbf{0.512} & 0.306 & 0.469 & -0.005 \\ 
 & $m=2.0$ & \textbf{0.451} & 0.241 & 0.379 & -0.002 \\ 
 & $m=2.2$ & \textbf{0.414} & 0.204 & 0.331 & -0.004 \\  \cline{2-6}
 $T=2000$ & $m=1.5$ & \textbf{0.648} & 0.487 & 0.687 & 0.011 \\ 
 & $m=1.8$ & \textbf{0.527} & 0.357 & 0.548 & 0.010 \\ 
 & $m=2.0$ & \textbf{0.477} & 0.282 & 0.445 & 0.008 \\ 
 & $m=2.2$ & \textbf{0.453} & 0.251 & 0.378 & 0.007 \\ 
   \hline
\end{tabular}
\caption{Average FARI obtained by the fuzzy $C$-medoids clustering procedures.}
\label{fuzzycmedoidstable}
\end{table}

We have redone the simulations by considering heavy tails in the error distribution. Note that this property often arises in real time series, specially within the field of finance \cite{harvey2013dynamic, bernardi2017multiple, rachev2003handbook, mikosch2003modeling}. Specifically, we have simulated the innovations in Scenarios 1 2 and 3 from a multivariate $t$ distribution with 3 degrees of freedom. For the sake of simplicity, we give the results only for the fuzzy $C$-means-based approaches. They are depicted in Table \ref{fuzzycmeanstablet}.

\begin{table}
\centering
\begin{tabular}{cccccc}
  \hline
   &  & QCD-FCMn & W-FCMn & C-FCMn & F-FCMn \\  \hline
   \multicolumn{6}{l}{Scenario 1}  \\ \hline
   $T=100$ & $m=1.5$ & \textbf{0.969} & 0.381 & 0.922 & 0.315 \\ 
   & $m=1.8$ & \textbf{0.927} & 0.321 & 0.796 & 0.299 \\ 
   & $m=2.0$ & \textbf{0.886} & 0.282 & 0.709 & 0.286  \\ 
   & $m=2.2$ & \textbf{0.839} & 0.243 & 0.629 & 0.270  \\ \cline{2-6}
   $T=150$ & $m=1.5$ & \textbf{0.990} & 0.458 & 0.954 & 0.356 \\ 
   & $m=1.8$ & \textbf{0.962} & 0.407 & 0.852 & 0.339  \\ 
   & $m=2.0$ & \textbf{0.929} & 0.367 & 0.773 & 0.325  \\ 
   & $m=2.2$ & \textbf{0.889} & 0.330 & 0.697 & 0.310  \\ \cline{2-6}
   $T=200$ & $m=1.5$ & \textbf{0.994} & 0.544 & 0.965 & 0.373  \\ 
   & $m=1.8$ & \textbf{0.971} & 0.488 & 0.876 & 0.357  \\ 
   & $m=2.0$ & \textbf{0.943} & 0.443 & 0.802 & 0.344  \\ 
   & $m=2.2$ & \textbf{0.907} & 0.402 & 0.729 & 0.329  \\ \hline 
    \multicolumn{6}{l}{Scenario 2}  \\ \hline
   $T=300$ & $m=1.5$ & \textbf{0.899} & 0.163 & 0.471 & 0.336  \\ 
   & $m=1.8$ & \textbf{0.851} & 0.160 & 0.413 & 0.310  \\ 
   & $m=2.0$ & \textbf{0.807} & 0.156 & 0.369 & 0.292 \\ 
   & $m=2.2$ & \textbf{0.759} & 0.156 & 0.328 & 0.267  \\ \cline{2-6}
   $T=400$ & $m=1.5$ & \textbf{0.957} & 0.151 & 0.484 & 0.347  \\ 
   & $m=1.8$ & \textbf{0.918} & 0.150 & 0.433 & 0.337  \\ 
   & $m=2.0$ & \textbf{0.879} & 0.149 & 0.393 & 0.316  \\ 
   & $m=2.2$ & \textbf{0.833} & 0.149 & 0.350 & 0.294 \\ \cline{2-6}
   $T=500$ & $m=1.5$ & \textbf{0.986} & 0.152 & 0.505 & 0.397  \\ 
   & $m=1.8$ & \textbf{0.956} & 0.151 & 0.455 & 0.383  \\ 
   & $m=2.0$ & \textbf{0.922} & 0.149 & 0.415 & 0.360  \\ 
   & $m=2.2$ & \textbf{0.881} & 0.150 & 0.377 & 0.334  \\  \hline
    \multicolumn{6}{l}{Scenario 3}  \\ \hline
   $T=1000$ & $m=1.5$ & \textbf{0.723} & 0.002 & 0.233 & -0.001  \\ 
   & $m=1.8$ & \textbf{0.644} & 0.001 & 0.189 & 0  \\ 
   & $m=2.0$ & \textbf{0.590} & 0.001 & 0.165 & -0.001  \\ 
   & $m=2.2$ & \textbf{0.540} & 0.001 & 0.143 & -0.001  \\ \cline{2-6}
   $T=1500$ & $m=1.5$ & \textbf{0.828} & 0.001 & 0.163 & 0.009 \\ 
   & $m=1.8$ & \textbf{0.735} & 0.001 & 0.138 & 0.007  \\ 
   & $m=2.0$ & \textbf{0.668} & 0.001 & 0.120 & 0.006  \\ 
   & $m=2.2$ & \textbf{0.610} & 0.001 & 0.104 & 0.005  \\  \cline{2-6}
   $T=2000$ & $m=1.5$ & \textbf{0.881} & 0.002 & 0.111 & 0  \\ 
   & $m=1.8$ & \textbf{0.783} & 0.002 & 0.090 & 0  \\ 
   & $m=2.0$ & \textbf{0.716} & 0.002 & 0.085 & 0  \\ 
   & $m=2.2$ & \textbf{0.653} & 0.002 & 0.074 & 0.001 \\ 
   \hline
\end{tabular}
\caption{Average FARI obtained by the fuzzy $C$-means clustering procedures. Innovations were drawn from a multivariate $t$ distribution with 3 degrees of freedom.}
\label{fuzzycmeanstablet}
\end{table}

By comparing Table \ref{fuzzycmeanstable} with Table \ref{fuzzycmeanstablet}, one can state that QCD-FCMn did not suffer when some amount of fat tailedness was introduced in the error distribution. It inherits the robustness of quantile methods, which gives this technique a desirable stability against the distributional form of the error. On the contrary, W-FCMn and C-FCMn significantly decreased their efficacy, specially in Scenarios 2 and 3. These procedures rely mainly on measures of traditional correlation and cross-correlation, and these measures get less accurate under tail behaviour.

It is worth noting that some other indexes aside from the FARI have been computed for the simulations carried out throughout this section. Particularly, the fuzzy versions of the Rand Index, the Jaccard Index and the Fowlkes-Mallows Index. All of them lead to similar conclusions as the ones stated above. Thus, the results concerning the mentioned alternative indexes are not presented in this work, although they are available upon request. 

\subsubsection{Second assessment scheme}\label{subsubsectionsecondassessment}

In this section the proposed approaches and their competitors are evaluated in a different manner than that of Section \ref{subsubsectionfirstassessment}. Here, we considered scenarios with two well separated clusters consisting of five time series each and a single switching series arising from a different generating process. The corresponding scenarios and generating processes are described below. \\

\noindent \textbf{Scenario 4}. Fuzzy clustering of VARMA processes with switching series. \\

\noindent Cluster 1: VAR(1) 

\begin{equation*}
\begin{pmatrix}
X_{t,1}  \\
X_{t,2} 
\end{pmatrix} = 
\begin{pmatrix}
0 & 0.2 \\
0.2 & 0.2
\end{pmatrix} 
\begin{pmatrix}
X_{t-1,1}   \\
X_{t-1,2}
\end{pmatrix} +
\begin{pmatrix}
\epsilon_{t,1}   \\
\epsilon_{t,2}
\end{pmatrix},
\end{equation*}

\noindent Cluster 2: VMA(1)
\begin{equation*}
\begin{pmatrix}
X_{t,1}  \\
X_{t,2} 
\end{pmatrix} = 
\begin{pmatrix}
-0.4 & -0.4   \\
-0.4 & -0.2 
\end{pmatrix} 
\begin{pmatrix}
X_{t-1,1}   \\
X_{t-1,2}
\end{pmatrix} +
\begin{pmatrix}
\epsilon_{t,1}   \\
\epsilon_{t,2}
\end{pmatrix},
\end{equation*}

\noindent Switching series: VARMA(1,1)
\begin{equation*}
	\begin{pmatrix}
		X_{t,1}  \\
		X_{t,2} 
	\end{pmatrix} = 
	\begin{pmatrix}
		0 & 0.2   \\
		0.2 & 0.2 
	\end{pmatrix} 
	\begin{pmatrix}
		X_{t-1,1}   \\
		X_{t-1,2}
	\end{pmatrix} + 
	\begin{pmatrix}
		-0.4 & -0.4   \\
		-0.4 & -0.2 
		\end{pmatrix} 
	\begin{pmatrix}
		\epsilon_{t-1,1}   \\
		\epsilon_{t-1,2}
	\end{pmatrix} +  
	\begin{pmatrix}
		\epsilon_{t,1}   \\
		\epsilon_{t,2}
	\end{pmatrix}.
\end{equation*} \\

%

\noindent \textbf{Scenario 5}. Fuzzy clustering of nonlinear processes with switching series. \\

\noindent Cluster 1: NLVMA (nonlinear VMA process)
\begin{equation*}
\begin{pmatrix}
X_{t,1}  \\
X_{t,2} 
\end{pmatrix} = 
\begin{pmatrix}
0.1\epsilon_{t-1,1}+0.6\epsilon_{t-1,2} ^2 + \epsilon_{t,1}\\ 
0.1\epsilon_{t-1,2}+0.6\epsilon_{t-1,1} ^2 + \epsilon_{t,2}
\end{pmatrix} 
+
\begin{pmatrix}
\epsilon_{t,1}   \\
\epsilon_{t,2}
\end{pmatrix},
\end{equation*}

\noindent Cluster 2: NLVMA
\begin{equation*}
\begin{pmatrix}
X_{t,1}  \\
X_{t,2} 
\end{pmatrix} = 
\begin{pmatrix}
-0.1\epsilon_{t-1,1}-0.6\epsilon_{t-1,2} ^2 + \epsilon_{t,1}\\ 
-0.1\epsilon_{t-1,2}-0.6\epsilon_{t-1,1} ^2 + \epsilon_{t,2}
\end{pmatrix} 
+
\begin{pmatrix}
\epsilon_{t,1}   \\
\epsilon_{t,2}
\end{pmatrix},
\end{equation*}

\noindent Switching series: a bivariate white noise process. 
\\

\noindent \textbf{Scenario 6}. Fuzzy clustering of dynamic conditional correlation processes with switching series. Consider the GARCH models and the correlation between the standardized shocks, $\rho_t$, in Scenario 3. \\

\noindent Cluster 1: $\rho_t=0.9I_{\{t\leq(T/2)\}}-0.3I_{\{t>(T/2)\}}$, \\ \\
Cluster 2: $\rho_t=-0.9I_{\{t\leq(T/2)\}}+0.3I_{\{t>(T/2)\}}$, \\\\
Switching series: $\rho_t=0$. \\

The error vector in Scenarios 4 and 5 follows a bivariate standard Gaussian distribution. 

Nonlinear ARMA processes have attracted a great deal of attention in the univariate framework \cite{pham2010hybrid, lu2003nonlinear, yassin2016binary, awwad2008nonlinear, boynton2013application}, proving themselves useful in several application fields. Particularly, nonlinear MA models have given rise to specific works due to the difficulty in their estimation \cite{karakucs2016bayesian, delgado2012parameters}. In a multivariate context, some works have also dealt with the topic of nonlinear VARMA models \cite{khan2015advances, weise1999asymmetric}.

Note that Scenarios 4, 5 and 6 have been designed in a way that the switching series is expected to lay ``in the middle'' of both clusters. In other words, a distance measure aimed to discriminate among generating processes should be able to produce very similar distance values from the switching series to a series from Cluster 1 and Cluster 2 indistinctly. In order to see if the proposed metric $d_{QCD}$ verifies this fact, we performed a metric two-dimensional scaling (2DS) based on the pairwise QCD-based dissimilarity matrix. The 2DS performs in the following way. Given a distance matrix $\bm D=(D_{ij})$, $i,j=1,\ldots,n$, it finds the set of points $\{(a_i, b_i), i = 1,\ldots, n\}$ such that the stress function

\begin{equation}\label{stress}
	\sqrt{\frac{\sum_{i \ne j=1}^{n}(\norm{(a_i, b_i)-(a_j, b_j)}-D_{ij})^2}{\sum_{i \ne j = 1}^{n}D_{ij}^2}}
\end{equation}

\noindent is minimized. Thus, the goal is to represent the distances $D_{ij}$ in terms of Euclidean distances into a 2-dimensional space so that the smaller the value of the stress function, the better the 2DS representation. The corresponding 2-D plot gives usually a good visual representation of how the elements are located with respect to each other according to the considered distance.

\begin{figure}
\centering
\includegraphics[width=1\textwidth]{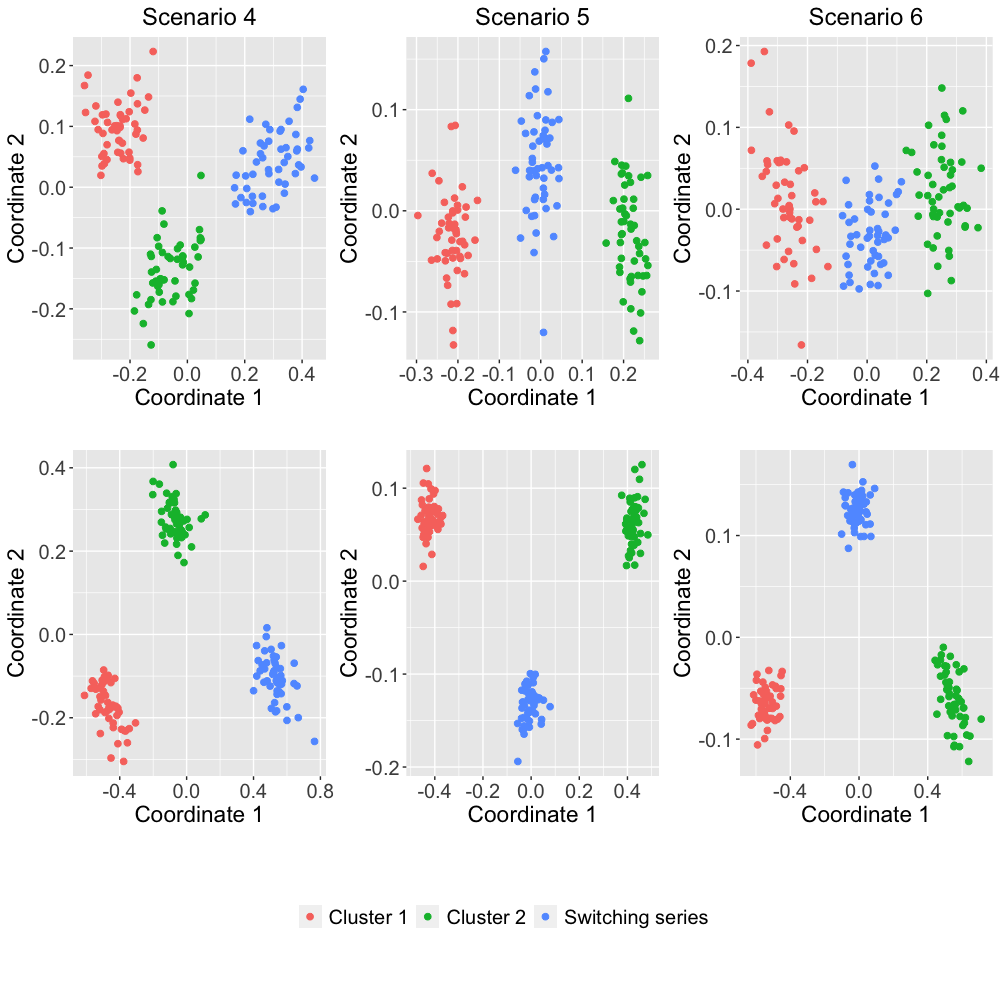}
\caption{Two-dimensional scaling planes based on QCD distances between simulated series in Scenarios 4, 5 and 6. The series length is $T=500$ in the top panels and $T=2000$ in the bottom panels.}
\label{mds}
\end{figure}

In order to visualize the corresponding 2-D graphs for each scenario, we simulated 50 MTS of length $T=500$ and $T=2000$ from the series defining the clusters and the switching series. The 2DS was carried out for each set of 150 MTS and value of the series length. The corresponding points in the new coordinate space are depicted in Figure \ref{mds}, where each set of points has been coloured according to the underlying generating process. The top panels show the 2DS for $T=500$ and the bottom panels, for $T=2000$.

With the aim of assessing the quality of the embedding, we computed the $R^2$ value to determine what proportion of variance of the scaled data can be accounted for the 2DS procedure. We obtained the values 0.667 (Scenario 4), 0.507 (Scenario 5) and 0.604 (Scenario 6) for the small sample sizes and 0.866 (Scenario 4), 0.788 (Scenario 5) and 0.852 (Scenario 6) for the large sample sizes. It is worth remarking that values above 0.6 are considered to provide an acceptable scaling procedure, whereas values above 0.8 mean a very good fit \cite{hair2009multivariate}. Although the values for the nonlinear case are lower than those for the remaining scenarios, they are quite close to the mentioned thresholds. Thus, generally, the plots in Figure \ref{mds} provide an accurate picture about the distance $d_{QCD}$ between the different underlying processes. 

The reduced bivariate spaces in Figure \ref{mds} show three compact, well-separated groups defined by the series of Cluster 1, Cluster 2, and the set of switching series. In all cases, the group of switching series (blue colour) is prone to position at an intermediate place between the remaining both clusters.This is a great trait exhibited by the distance $d_{QCD}$, since, as remarked previously, Scenarios 4, 5 and 6 were chosen so that the switching series is in the middle of Clusters 1 and 2 in terms of generating processes. Note that, for a given value of the series length, the most challenging scenario for the proposed clustering algorithms seems to be Scenario 6, since the 2DS plot produces near overlapping clusters when $T=500$. As expected, when increasing the series length, the QCD-based features are more accurately estimated and the groups become more distant from one another. In summary, Figure \ref{mds} uncovers that the QCD-based distance should be capable of discriminating between the original clusters and the switching series in Scenarios 4, 5 and 6, thus being a good candidate for the second assessment design. 

We considered again different values for the series length, namely, $T\in \{200, 400, 600\}$ in Scenario 4, $T\in \{300, 600, 900\}$ in Scenario 5 and $T\in \{500, 1000, 1500\}$ in Scenario 6. The values taken into account for the fuzziness parameter $m$ were the same as in Section \ref{subsubsectionfirstassessment}. Again, 200 simulations were performed, and the fuzzy $C$-means and fuzzy $C$-medoids versions of the procedures were applied. The same hyperparameters as in Section \ref{subsubsectionfirstassessment} were taken into consideration to obtain the QCD-based features and those concerning the alternative procedures. This time, the number of clusters was set to $C=2$. In this second evaluation scheme, we assessed the clustering methods by means of the frequency with which the five series from Cluster 1 grouped together in one group, the five series series from Cluster 2 clustered together in another group, and the switching series had a relatively high membership degree with regards to both groups. To that end, we had to define a cutoff point in order to decide when a given realization was assigned to a specific cluster. We decided to use the cutoff value of 0.7, so that the $i$-th MTS was placed into the $c$-th cluster if $u_{ic}>0.7$. This cutoff value has already been considered in some works \cite{d2012wavelets, d2009autocorrelation, maharaj2011fuzzy}. A discussion about the reasoning for choosing this membership degree constraint can be seen in \cite{maharaj2011fuzzy}. In this way, the switching series was considered to concurrently pertain to both clusters if its membership degrees were both below 0.7. 

\subsubsection*{Results}

The average success rates attained by the $C$-means techniques according to the previous criteria are presented in Table \ref{fuzzycmeanstableswitching}. The best approach was QCD-FCMn, achieving the best average scores in all the considered configurations excluding Scenario 4 with $m=1.5$ and $T=600$, where it was outplayed by F-FCMn. As expected, its results improved when increasing $T$. It also got better for larger values of $m$ in the three scenarios. This is owing to the fact that, for the smaller values of $m$, QCD-FCMn returns a quite crisp partition so that the switching series is located a lot of times in one cluster with high membership. Indeed this is the cause that QCD-FCMn attained poor classification rates for $m=1.5$ whatever the value of the series length in Scenario 4. However, when $m$ becomes larger, the boundaries between clusters become blurrier and the switching series is simultaneously located in both clusters. Nevertheless, this gives rise to a different type of classification error, since, for larger values of $m$, frequently some of the non switching series display similar membership values in both clusters (e.g. 0.55 and 0.45). This trade-off situation concerning Scenario 4 for QCD-FCMn gets fixed as the value of $T$ increases. 

\begin{table}
\centering
\begin{tabular}{cccccc}
  \hline
 &  & QCD-FCMn & W-FCMn & C-FCMn & F-FCMn \\  \hline
  \multicolumn{6}{l}{Scenario 4} \\ \hline
 $T=200$ & $m=1.5$ & \textbf{0.070} & 0.005 & 0 & 0.015 \\ 
   & $m=1.8$ &  \textbf{0.130} & 0.010 & 0.005 & 0.015 \\ 
   & $m=2.0$ &  \textbf{0.115} & 0.015 & 0.025 & 0.015 \\ 
   & $m=2.2$ &  \textbf{0.065} & 0.015 & 0.030 & 0.010 \\ \cline{2-6} 
   $T=400$ & $m=1.5$ &  \textbf{0.070} & 0 & 0 & 0.065 \\ 
   & $m=1.8$ &  \textbf{0.315} & 0 & 0 & 0.110 \\ 
   & $m=2.0$ &  \textbf{0.495} & 0.025 & 0.010 & 0.125 \\ 
   & $m=2.2$ &  \textbf{0.455} & 0.075 & 0.070 & 0.085 \\  \cline{2-6} 
   $T=600$ & $m=1.5$ & 0.055 & 0 & 0 &  \textbf{0.245} \\ 
   & $m=1.8$ &  \textbf{0.440} & 0 & 0 & 0.390 \\ 
   & $m=2.0$ &  \textbf{0.610} & 0.005 & 0.015 & 0.395 \\ 
   & $m=2.2$ &  \textbf{0.725} & 0.045 & 0.115 & 0.410 \\  \hline
   \multicolumn{6}{l}{Scenario 5} \\ \hline
   $T=300$ & $m=1.5$ &  \textbf{0.435} & 0 & 0.025 & 0 \\ 
   & $m=1.8$ &  \textbf{0.610} & 0 & 0.005 & 0 \\ 
   & $m=2.0$ &  \textbf{0.700} & 0 & 0.005 & 0 \\ 
   & $m=2.2$ &  \textbf{0.650} & 0 & 0.005 & 0 \\ \cline{2-6}
  $T=600$ & $m=1.5$ &  \textbf{0.520} & 0 & 0.090 & 0 \\ 
   & $m=1.8$ &  \textbf{0.810} & 0 & 0.085 & 0 \\ 
   & $m=2.0$ &  \textbf{0.895} & 0 & 0.065 & 0 \\ 
   & $m=2.2$ &  \textbf{0.945} & 0 & 0.040 & 0 \\  \cline{2-6}
  $T=900$ & $m=1.5$ &  \textbf{0.690} & 0 & 0.150 & 0 \\ 
   & $m=1.8$ &  \textbf{0.920} & 0.010 & 0.220 & 0 \\ 
   & $m=2.0$ &  \textbf{0.980} & 0.005 & 0.170 & 0 \\ 
   & $m=2.2$ &  \textbf{0.990} & 0.005 & 0.105 & 0 \\  \hline 
   \multicolumn{6}{l}{Scenario 6} \\ \hline
   $T=500$ & $m=1.5$ &  \textbf{0.370} & 0.280 & 0.330 & 0 \\ 
   & $m=1.8$ &  \textbf{0.645} & 0.450 & 0.560 & 0 \\ 
   & $m=2.0$ &  \textbf{0.735} & 0.515 & 0.670 & 0 \\ 
   & $m=2.2$ &  \textbf{0.820} & 0.500 & 0.695 & 0 \\  \cline{2-6}
   $T=1000$ & $m=1.5$ &  \textbf{0.500} & 0.335 & 0.395 & 0 \\ 
   & $m=1.8$ &  \textbf{0.790} & 0.595 & 0.595 & 0 \\ 
   & $m=2.0$ &  \textbf{0.890} & 0.675 & 0.725 & 0 \\ 
   & $m=2.2$ &  \textbf{0.955} & 0.735 & 0.840 & 0 \\  \cline{2-6}
   $T=1500$ & $m=1.5$ &  \textbf{0.550} & 0.415 & 0.425 & 0 \\ 
   & $m=1.8$ &  \textbf{0.860} & 0.680 & 0.730 & 0 \\ 
   & $m=2.0$ &  \textbf{0.940} & 0.785 & 0.850 & 0 \\ 
   & $m=2.2$ &  \textbf{0.980} & 0.840 & 0.920 & 0 \\ 
   \hline
\end{tabular}
	\caption{Average success rates obtained by the fuzzy $C$-means clustering procedures.}
\label{fuzzycmeanstableswitching}
\end{table}

The approaches W-FCMn and C-FCMn only reached satisfactory success rates in Scenario 6, whereas they completely failed in Scenarios 4 and 5. This seems surprising, since at least C-FCMn was expected to perform well when dealing with Scenario 4 as the corresponding metric should be able to distinguish between linear processes. We ran this procedure for values of $m$ greater than $2.2$ in Scenario 4 and found out that the corresponding distance was able to get higher success rates for some of those values (but less than those associated with QCD-FCMn). The reason was that, for C-FCMn, the value $m=2.2$ still means a very crisp partition so its failures are attributable to the switching series. This fact highlights the paramount importance of the fuzziness parameter $m$ when evaluating the strategies through a cutoff point.

Table \ref{fuzzycmedoidstableswitching} contains the average frequencies of correct classification for the fuzzy $C$-medoids variants of the methods. Generally speaking, QCD-FCMd showed approximately the same performance as QCD-FCMn in the three scenarios. W-FCMd and C-FCMd somewhat improved their scores in Scenario 4 but worsened it in Scenarios 5 and 6. F-FCMd achieved better scores than F-FCMd in Scenario 4. As in Table \ref{fuzzycmeanstableswitching}, the results appear to be highly influenced by the values of the fuzziness coefficient $m$. 

\begin{table}
\centering
\begin{tabular}{cccccc}
  \hline
   &  & QCD-FCMd & W-FCMd & C-FCMd & F-FCMd \\  \hline
    \multicolumn{6}{l}{Scenario 4} \\ \hline
  $T=200$ & $m=1.5$ & \textbf{0.080} & 0.010 & 0.010 & 0.020 \\ 
  & $m=1.8$ & \textbf{0.085} & 0.025 & 0.020 & 0.015 \\ 
  & $m=2.0$ & \textbf{0.075} & 0.030 & 0.075 & 0.020 \\ 
  & $m=2.2$ & 0.045 & 0.020 & \textbf{0.060} & 0.005 \\ \cline{2-6}
  $T=400$ & $m=1.5$ & 0.135 & 0 & 0.005 & \textbf{0.180} \\ 
  & $m=1.8$ & \textbf{0.365} & 0.025 & 0.035 & 0.180 \\ 
  & $m=2.0$ & \textbf{0.395} & 0.065 & 0.085 & 0.130 \\ 
  & $m=2.2$ & \textbf{0.360} & 0.140 & 0.240 & 0.110 \\ \cline{2-6}
  $T=600$ & $m=1.5$ & 0.130 & 0 & 0 & \textbf{0.415} \\ 
  & $m=1.8$ & 0.415 & 0.005 & 0.020 & \textbf{0.450} \\ 
  & $m=2.0$ & \textbf{0.570} & 0.070 & 0.120 & 0.440 \\ 
  & $m=2.2$ & \textbf{0.655} & 0.215 & 0.285 & 0.355 \\ \hline 
     \multicolumn{6}{l}{Scenario 5} \\ \hline
  $T=300$ & $m=1.5$ & \textbf{0.425} & 0 & 0.005 & 0 \\
  & $m=1.8$ & \textbf{0.540} & 0 & 0 & 0 \\ 
  & $m=2.0$ & \textbf{0.620} & 0 & 0 & 0 \\ 
  & $m=2.2$ & \textbf{0.540} & 0 & 0 & 0 \\ \cline{2-6}
  $T=600$ & $m=1.5$ & \textbf{0.600} & 0 & 0.040 & 0 \\ 
  & $m=1.8$ & \textbf{0.820} & 0 & 0.025 & 0 \\ 
  & $m=2.0$ & \textbf{0.895} & 0 & 0.015 & 0 \\ 
  & $m=2.2$ & \textbf{0.930} & 0 & 0.005 & 0 \\ \cline{2-6}
  $T=900$ & $m=1.5$ & \textbf{0.700} & 0 & 0.145 & 0 \\ 
  & $m=1.8$ & \textbf{0.910} & 0.005 & 0.095 & 0 \\ 
  & $m=2.0$ & \textbf{0.960} & 0 & 0.035 & 0 \\ 
  & $m=2.2$ & \textbf{0.975} & 0 & 0.015 & 0 \\ \hline
     \multicolumn{6}{l}{Scenario 6} \\ \hline
  $T=500$ & $m=1.5$ & \textbf{0.400} & 0.315 & 0.360 & 0 \\ 
  & $m=1.8$ & \textbf{0.650} & 0.440 & 0.535 & 0 \\ 
  & $m=2.0$ & \textbf{0.735} & 0.455 & 0.580 & 0 \\ 
  & $m=2.2$ & \textbf{0.800} & 0.455 & 0.635 & 0 \\ \cline{2-6}
  $T=1000$ & $m=1.5$ & \textbf{0.545} & 0.375 & 0.405 & 0 \\ 
  & $m=1.8$ & \textbf{0.795} & 0.510 & 0.605 & 0 \\ 
  & $m=2.0$ & \textbf{0.880} & 0.585 & 0.685 & 0 \\ 
  & $m=2.2$ & \textbf{0.950} & 0.655 & 0.760 & 0 \\ \cline{2-6}
  $T=1500$ & $m=1.5$ & \textbf{0.610} & 0.425 & 0.430 & 0 \\ 
  & $m=1.8$ & \textbf{0.840} & 0.660 & 0.685 & 0 \\ 
  & $m=2.0$ & \textbf{0.925} & 0.775 & 0.800 & 0 \\ 
  & $m=2.2$ & \textbf{0.980} & 0.820 & 0.860 & 0 \\ 
   \hline
\end{tabular}
\caption{Average success rates obtained by the fuzzy $C$-medoids clustering procedures.}
\label{fuzzycmedoidstableswitching}
\end{table}

\begin{figure}
	\centering
	\includegraphics[width=1\textwidth]{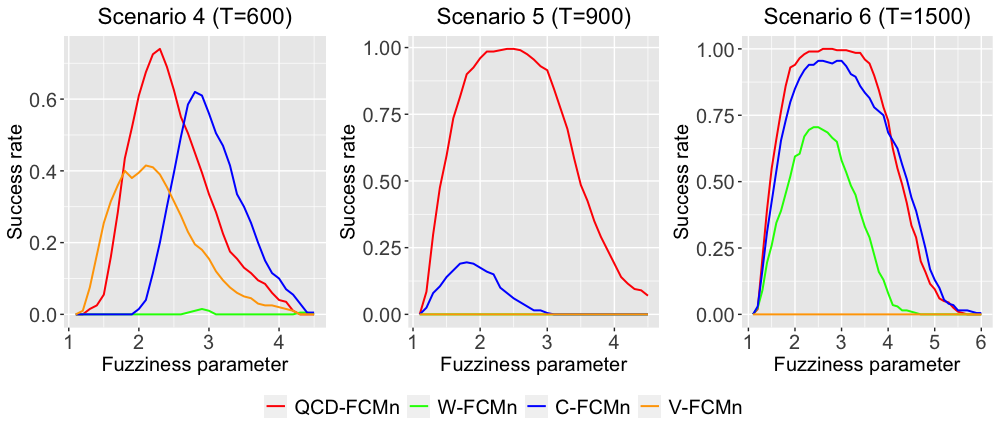}
	\caption{Average rates of correct classification for the fuzzy $C$-means procedures as a function of $m$.}
	\label{curvesmcmeans}
\end{figure}

In order to complement the evaluation results of Tables \ref{fuzzycmeanstableswitching} and \ref{fuzzycmedoidstableswitching}, we considered the largest values of the series length (600, 900 and 1500) and a grid of values for $m$ equispaced between 1 and a value large enough so that all the methods achieve near-zero rates of correct classification. Simulations were performed in the same way as before. The corresponding curves of frequencies of correct classification as a function of $m$ for the four fuzzy $C$-means methods are shown in Figure \ref{curvesmcmeans}. These curves give a much more complete picture about the discriminatory capability of the approaches over the whole range of values for the fuzziness parameter. As stated before, although values of $m$ above 2.5 or 3 are not commonly used in practice, they can be taken into consideration in the assessment mechanism in order to get a fair comparison between the considered approaches. Given the graphs in Figure \ref{curvesmcmeans}, a reasonable measure of performance is the area under the fuzziness curve (AUFC). The corresponding quantities are given in the first part of Table \ref{tablem}. QCD-FCMn was the approach associated with the greatest value of AUFC in the three scenarios. It clearly outmatched the remaining strategies in Scenarios 1 and 3, and got a slightly better value than C-FCMn in Scenario 3. This latter approach was by far the second best performing one. 

\begin{table}
	\centering 
	\begin{tabular}{cccccc} \hline 
		&            & QCD-FCMn & W-FCMn & C-FCMn & F-FCMn \\ \hline
		Fuzzy                & Scenario 1 & \textbf{0.9205}   & 0.005  & 0.6943 & 0.5845 \\
	$C$-means	& Scenario 2 & \textbf{2.0855}   & 0      & 0.1925 & 0      \\
		& Scenario 3 & \textbf{2.7825}   & 1.2710 & 2.6483 & 0      \\ \hline 
		Fuzzy             & Scenario 1 &     \textbf{0.7620}     &    0.0002    &   0.5740     &    0.6290    \\
	$C$-medoids 	& Scenario 2 &    \textbf{1.9908}      &   0     &     0.0975   &    0    \\
		& Scenario 3 &       \textbf{2.5923}   &    0.9825    &   2.2283     &   0     \\ \hline 
		Fuzzy $C$-means  & Scenario 1 &   \textbf{1.1568}       &  0.0215        &    0.0495    &     0.1635   \\
	(heavy tails)	& Scenario 2 &    \textbf{3.7380}      &   0     &   0     &    0    \\
		& Scenario 3 &    \textbf{2.2058}     &    0    &   0   &  0   \\   \hline 
	\end{tabular}
\caption{Area under the curves of Figures \ref{curvesmcmeans}, \ref{curvesmcmedoids} and \ref{curvesmcmeanst}.}
\label{tablem}
\end{table}

\begin{figure}
	\centering
	\includegraphics[width=1\textwidth]{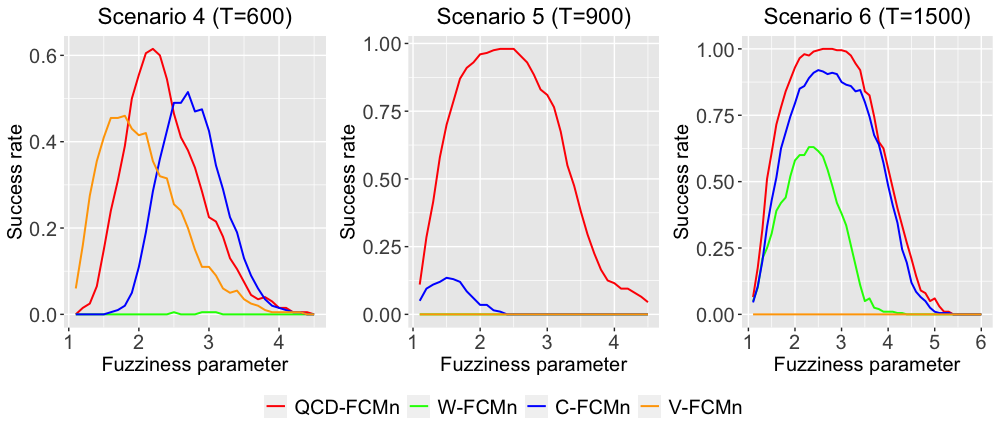}
	\caption{Average rates of correct classification for the fuzzy $C$-medoids procedures as a function of $m$.}
	\label{curvesmcmedoids}
\end{figure}

Figure \ref{curvesmcmedoids} shows the fuzziness curves for the fuzzy $C$-medoids strategies. The situation is rather similar to that of Figure \ref{curvesmcmedoids}. On the whole, there is a tiny decline in the performance of all techniques concerning most values of $m$. The respective values of the AUFC are presented in the second part of Table \ref{tablem}. Indeed, there is a decrease in AUFC for all the methods in comparison with the fuzzy $C$-means setting except for F-FCMn in Scenario 1, which slightly improved its score. Again, QCD-FCMn reached the best overall results among the four analysed algorithms. 

\begin{figure}
	\centering
	\includegraphics[width=1\textwidth]{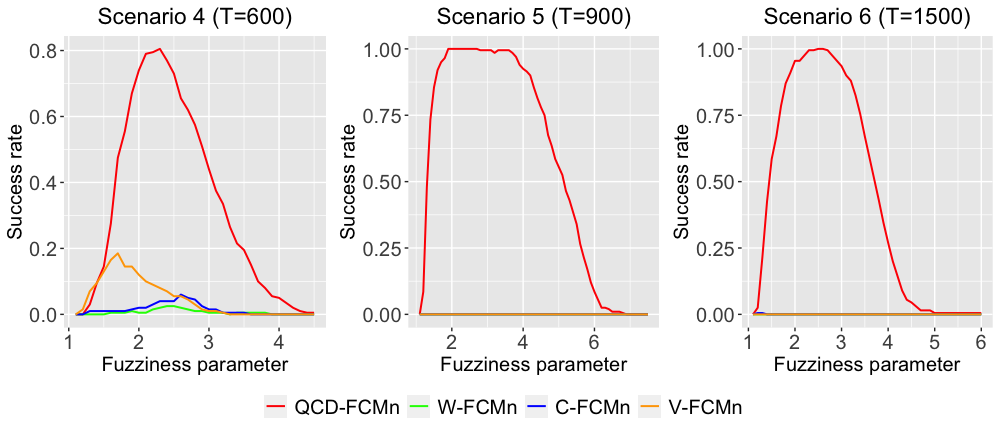}
	\caption{Average rates of correct classification for the fuzzy $C$-means procedures as a function of $m$. Innovations were drawn from a multivariate t distribution with 3 degrees of freedom.}
	\label{curvesmcmeanst}
\end{figure}

As in Section \ref{subsubsectionfirstassessment}, the simulations concerning Scenarios 4, 5 and 6 were repeated by taking into account a multivariate $t$ distribution with 3 degrees of freedom for the innovations. The respective fuzziness curves for the fuzzy $C$-means methods are displayed in Figure \ref{curvesmcmeanst}, whereas the corresponding values for the AUFC are shown in the last part of Table \ref{tablem}. Clearly, QCD-FCMn is the only approach capable of performing  an effective classification under these circumstances, exhibiting again a substantial robustness against the departure from normality in the error distribution. The remaining procedures got a poor rate of correct classification in Scenario 1 and zero in Scenarios 2 and 3.

It is worth highlighting that, in all the previous analyses, the fuzzy $C$-means and the fuzzy $C$-medoids algorithms were performed over many random assignments in order to avoid the issues of local optima.

\section{Application}\label{sectionapplication}

This section is devoted to show the application of the proposed clustering procedures.

\subsection{Fuzzy clustering of the top 20 companies in the S\&P 500 index}\label{subsectionsp500}

Time series clustering has proven very useful when dealing with financial time series. More often than not, the aim is to group different assets or companies according to how they behave over a certain period of time. This way, the clustering solution can represent groups sharing similar risk profiles, management behaviour, or even future expected returns. There are a broad variety of works coping with clustering of financial time series \cite{kou2014evaluation, durante2014clustering, pattarin2004clustering, dias2015clustering, bastos2014clustering, d2013clustering}. Here we present how one of the proposed fuzzy clustering approaches can be applied to group the most important companies in the US. We want to remark that the following analyses are not aimed at giving financial advice nor deriving economical implications, but at illustrating the suitability of the designed technique to recognize homogeneous groups with similar stochastic dependence patterns and to what extent each company pertains to each one of the identified groups. 

The data we are going to use was taken from the finance section of the Yahoo website\footnote{https://es.finance.yahoo.com}. It contains daily stock returns and trading volume of the current top 20 companies of the S\&P 500 index according to market capitalization. The sample period spans from 6th July 2015 to 7th February 2018, thus resulting serial realizations of length $T=655$. The S\&P 500 is a stock market index that tracks the stocks of 500 large-cap U.S. companies. The top 20 contains some of the most important companies in the world, as Apple, Google, Facebook or Berkshire Hathaway. 

It is important to highlight that the relationship between price and volume has been extensively analyzed in the literature \cite{karpoff1987relation, campbell1993trading, gebka2013causality} and constitutes itself a topic of great financial interest. Prices and trading volume are known to exhibit some empirical linkages over the fluctuations of stock markets. Thus, it is reasonable to describe each of the considered companies by means of these two quantities. Our goal is to analyze the joint behaviour of prices and volume in order to perform fuzzy clustering. Thus, we assume that two companies behave similarly if the corresponding bivariate time series exhibit similar dependence structures. 

\begin{figure}
	\centering
	\includegraphics[width=1\textwidth]{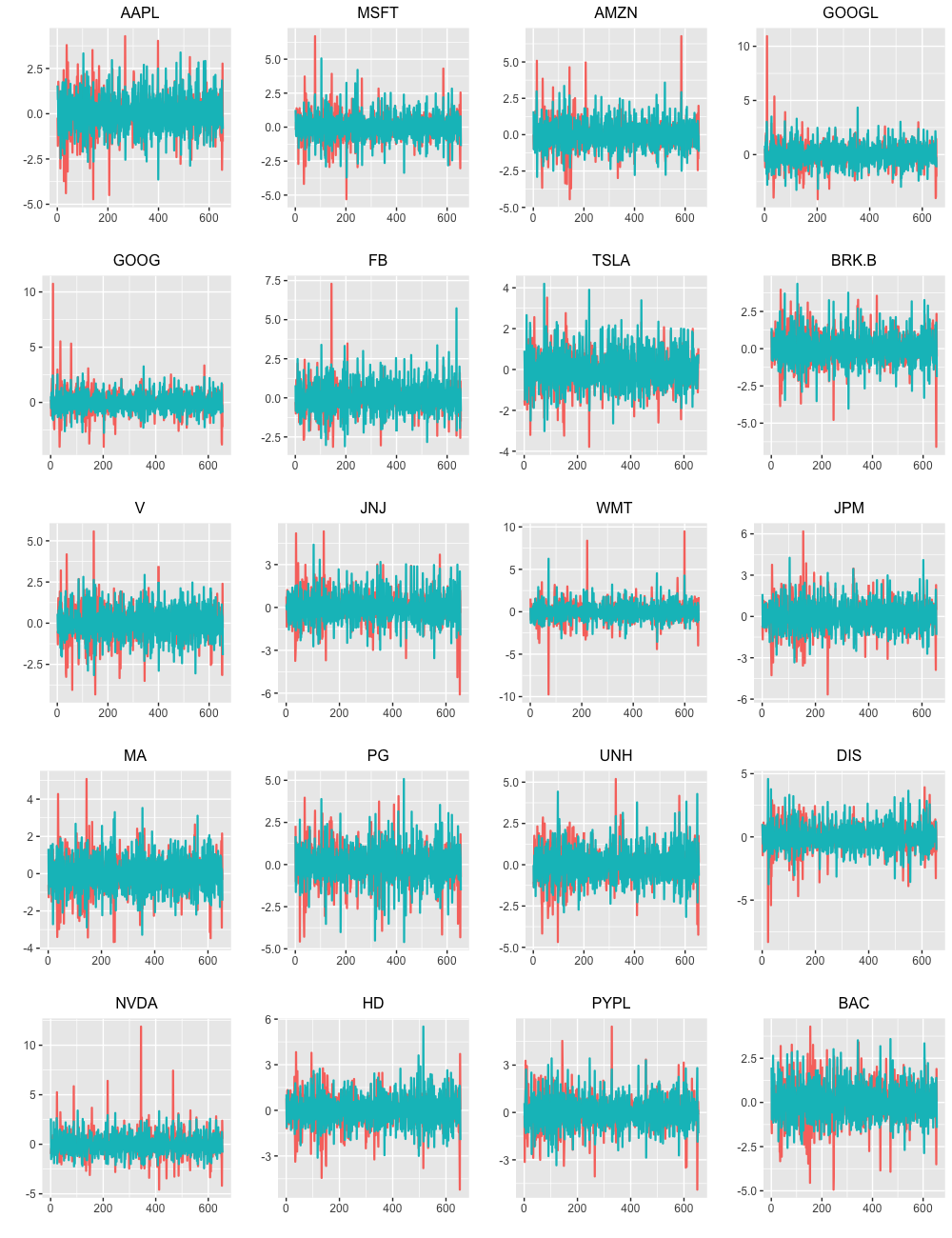}
	\caption{Daily returns (red colour) and change in volume (blue colour) of the top 20 companies in the S\&P 500 index.}
	\label{top20}
\end{figure}

It can be observed that both the UTS of prices and trading volume are non-stationary in mean. Thus, all UTS are transformed by taking the first differences of the natural logarithm of the original values. This way, prices give rise to stock returns, and volume to what we call change in volume. Finally, all UTS are normalized to have zero mean and unit variance. The resulting MTS are shown in Figure \ref{top20}, where the returns and the change in volume are displayed through the red and blue colour, respectively. Overall, plots in Figure \ref{top20} exhibit common traits of financial time series. There is a substantial degree of heteroskedasticity in both prices and change in volume. In addition, both quantities exhibit the so-called phenomenon of volatility clustering: large values (positive or negative) tend to group together, resulting in a marked persistence. These special properties of financial time series, usually referred to as stylized facts, are generally accounted for by modelling the series by way of multivariate GARCH-type models, for instance the dynamic conditional correlation models of Scenarios 3 and 6. It is worth remembering that the devised fuzzy clustering algorithms demonstrated their efficacy when coping with this type of models, specially when the error distribution possesses some amount of fat-tailedness. This property also relates to the stylized facts \cite{harvey2013dynamic, schmitt2013non, bradley2003financial}. Thus, given the high capability of QCD-FCMn and QCD-FCMd to discriminate between conditional heteroskedastic models, it is expected that both methods can provide a meaningful fuzzy partition determining groups of companies following a similar behavioural pattern.

\begin{figure}
	\centering
	\includegraphics[width=1\textwidth]{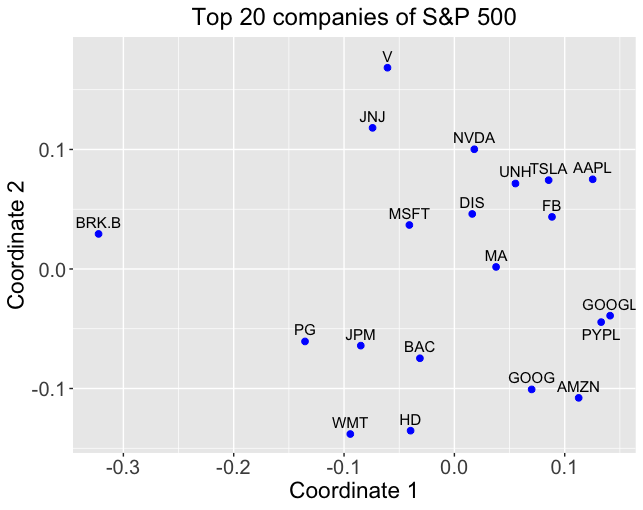}
	\caption{Two-dimensional scaling plane based on the QCD distances for the daily returns and change in volume of the top 20 companies in the S\&P 500 index.}
	\label{mdstop20}
\end{figure}

As a preliminary exploratory step, we performed a 2DS based on the pairwise QCD-dissimilarity matrix. That way, a projection of the companies on a two-dimensional plane preserving the original distances as well as possible is available. The location of the top 20 companies in the transformed space is displayed in Figure \ref{mdstop20}. The $R^2$ value is 0.7251. Thus, the scatter plot in Figure \ref{mdstop20} can be considered an acceptable representation of the underlying distance configuration \cite{hair2009multivariate}.

Overall, Figure \ref{mdstop20} suggests that the grouping of the top 20 companies calls for a fuzzy partition, since the points in the new coordinate space do not appear to be split in clear, well-separated, nonoverlapping groups. Hence, a fuzzy clustering algorithm is expected to give more meaningful insights into the distribution of the corporations than a hard clustering one, where each company would be allowed to pertain to only one cluster. By observing Figure \ref{mdstop20}, one could conclude the existence of 5 or 6 overlapping clusters. There is a cluster including the three technological giants Google Alphabet Class A (GOOGL), Google Alphabet Class C (GOOG) and Amazon (AMZN), along with the company PayPal (PYPL). The remaining tech giants Apple (AAPL), Facebook (FB) and Microsoft (MSFT) appear to form another group with some other businesses as Tesla (TSLA) and Walt Disney (DIS). The rest of the firms are more spread out. Berkshire Hathaway (BRK.B) seems to be isolated, well separated from the remaining companies, and it could be tought of as an atypical corporation. Johnson \& Johnson (JNJ) and Visa (V) are placed close to each other, so they could constitute another cluster. The five remaining organizations Procter \& Gamble (PG), JPMorgan Chase (JPM), Bank of America (BAC), Walmart (WMT) and Home Depot (HD) could comprise one or two different groups (PG-JPM-BAC and WMT-HD).

As the QCD-FCMn procedure achieved overall slightly better results than the QCD-FCMd algorithm in the simulation study carried out in Section \ref{sectionsimulationstudy}, the former was the method chosen for the application. Just as in the simulations, the metric $d_{QCD}$ was constructed by using the probability levels 0.1, 0.5 and 0.9. As it was already pointed out, and as we saw in Section \ref{sectionsimulationstudy}, the fuzziness parameter $m$ highly influences the quality of the obtained clustering partition. Thus, we decided to choose this parameter as well as the number of clusters, $C$, by way of a data driven approach. In order to do so, we took into consideration four different cluster validity indexes, the Xie-Beni Index (XBI) \cite{xie1991validity}, the Kwon Index (KI) \cite{kwon1998cluster} and the indexes proposed in \cite{tang2005improved} (TI) and in \cite{bensaid1996validity} (BI). Now we define those indexes according to the fuzzy approach based of features extracted from an MTS. Let $d_{ic}$ be the Euclidean distance between the element $\bm \varphi^{(i)}$ and the centroid of the $c$-th cluster, $\overline{\bm \varphi}^{(c)}$. The XBI is defined as

\begin{equation}\label{xbi}
	\text{XBI}(C,m)=\frac{\sum_{i=1}^{n}\sum_{c=1}^{C}u_{ic}^2d_{ic}^2}{n\min_{c\ne c'}\norm{\overline{\bm \varphi}^{(c)}-\overline{\bm \varphi}^{(c')}}^2}.
\end{equation}

Note that minimizing the numerator on \eqref{xbi}, which measures the compactness of the fuzzy partition, is precisely the goal of QCD-FCMn when $m=2$. The denominator in \eqref{xbi} measures the degree of separation between clusters. Thus, the index decreases with separation between clusters. Note that, for a given membership matrix $\bm U$ obtained as a solution of a clustering algorithm, XBI is a function of the fuzziness coefficient $m$.

The KI is an extension of the XBI which is aimed at penalizing the monotonically decreasing trend exhibited by the latter when $C$ becomes large. The KI is given by
\begin{equation}\label{ki}
\text{KI}(C,m)=\frac{\sum_{i=1}^{n}\sum_{c=1}^{C}u_{ic}^2d_{ic}^2+(1/C)\sum_{c=1}^{C}\norm{\overline{\bm \varphi}^{(c)}-\overline{\bm \varphi}}^2}{\min_{c\ne c'}\norm{\overline{\bm \varphi}^{(c)}-\overline{\bm \varphi}^{(c')}}^2},
\end{equation}
where $\overline{\bm \varphi}=\frac{1}{n}\sum_{i=1}^{n}{\bm \varphi}^{(i)}$. Note that the punishing term $(1/C)\sum_{c=1}^{C}\norm{\overline{\bm \varphi}^{(c)}-\overline{\bm \varphi}}^2$ increases substantially as the number of clusters is close to $n$.

Although KI allows to evaluate the quality of a fuzzy $C$-means procedure when $C\rightarrow n$, it becomes unstable or unpredictable as $m \rightarrow \infty$. The TI tries to solve this problem by adding a penalty function both in the numerator and the denominator. This index is defined as
\begin{equation}\label{ti}
\text{TI}(C,m)=\frac{\sum_{i=1}^{n}\sum_{c=1}^{C}u_{ic}^2d_{ic}^2+\frac{1}{C(C-1)}\sum_{c=1}^{C}\sum_{c'=1;c'\ne c}^{C}\norm{\overline{\bm \varphi}^{(c)}-\overline{\bm \varphi}^{(c')}}^2}{\min_{c\ne c'}\norm{\overline{\bm \varphi}^{(c)}-\overline{\bm \varphi}^{(c')}}^2+1/C}.
\end{equation}

The BI is given by
\begin{equation}\label{bi}
\text{BI}(C,m)=\sum_{c=1}^{C}\frac{\sum_{i=1}^{n}u_{ic}^2d_{ic}^2}{\sum_{i=1}^{n}u_{ic}\sum_{c'=1}^{C}\norm{\overline{\bm \varphi}^{(c)}-\overline{\bm \varphi}^{(c')}}^2}.
\end{equation}

The numerator in \eqref{bi} measures the compactness and the denominator accounts for the separation between clusters.

Note that, concerning all the previous defined indexes, smaller values mean better fuzzy partitions. Indeed, the minimum value of some of those indexes is frequently utilised to choose the optimal number of clusters \cite{lafuente2018robust, d2009autocorrelation} and the optimal value for the fuzziness parameter \cite{zhou2014fuzziness}. Here, our concern is to determine the combination of these two values which must be considered to optimize the clustering process. The goal is simple: starting from a grid of input parameters (several values for $C$ and $m$), we solve the minimization problem in \eqref{qcd_means} and select the values of $C$ and $m$ that lead to the minimum average value of the indexes XBI, KI, TI and BI. The process is carried out as follows. 
\begin{description}
\item[\textit{Step 1}.] Select the initial grid of values for the number of clusters, $C$, the fuzziness parameter, $m$; and the number of random inizializations for the fuzzy $C$-means clustering procedure, $R$.   
\item[\textit{Step 2}.] Solve the minimization problem in \eqref{qcd_means} for each pair $(C,m)$ in the grid by considering $R$ random inizializations for the centroids.  Store the resulting membership matrices and centroids.  
\item[\textit{Step 3}.] For each pair $(C,m)$ in the grid, obtain the $R$ values of \eqref{xbi}, \eqref{ki}, \eqref{ti} and \eqref{bi} and select the corresponding minimum values. The results of this step are 4 vectors of length $Cm$ storing the corresponding values of the indexes, $v_{\text{XBI}}$, $v_{\text{KI}}$, $v_{\text{TI}}$ and $v_{\text{BI}}$. 
\item[\textit{Step 4}.] Scale the vectors resulting from \textit{Step 3} so that they have zero mean and unit variance to get $v^s_{\text{XBI}}$, $v^s_{\text{KI}}$, $v^s_{\text{TI}}$ and $v^s _{\text{BI}}$.  
\item[\textit{Step 5}.] Obtain a vector giving the average of the scaled indexes for each pair $(C,m)$
\begin{equation*}
v_{\text{avg}}(C,m)=\frac{v^s_{\text{XBI}}(C,m)+v^s_{\text{KI}}(C,m)+v^s_{\text{TI}}(C,m)+v^s_{\text{BI}}(C,m)}{4}.
\end{equation*}
\item[\textit{Step 6}.] Return the pair $(C,m)$ giving rise to the minimum value of $v_{\text{avg}}$. 
 \end{description}
 
  Note that standardization in \textit{Step 4} is necessary to bring the four indexes to the same scale. 

We selected a grid between 1 and 10 (step size of 1) for the number of clusters $C$ and a grid between 1.1 and 3 (step size of 0.1) for the fuzziness parameter $m$ and applied the previous selection procedure by taking into account the QCD-FCMn algorithm. The optimal $(C,m)$ pair was $(6, 1.9)$. The value $C=6$ seems consistent with the plot in Figure \ref{mdstop20}.

Table \ref{tablefuzzy} shows the membership degrees obtained by taken into consideration the previous values of $C$ and $m$. For each single corporation, the entries in bold enhance the highest membership degrees, i.e, the cluster assignment from a crisp perspective.

\begin{table}
\centering
\begin{tabular}{ccccccc}
  \hline
 Company & $C_1$ & $C_2$ & $C_3$ & $C_4$ & $C_5$ & $C_6$ \\ 
  \hline
AAPL & 0.115 & 0.213 & 0.016 & \textbf{0.516} & 0.090 & 0.050 \\ 
  MSFT & 0.147 & 0.068 & 0.060 & \textbf{0.494} & 0.137 & 0.095 \\ 
  AMZN & \textbf{0.891} & 0.022 & 0.006 & 0.041 & 0.012 & 0.028 \\ 
  GOOGL & \textbf{0.757} & 0.039 & 0.013 & 0.127 & 0.028 & 0.036 \\ 
  GOOG & \textbf{0.923} & 0.012 & 0.005 & 0.032 & 0.009 & 0.019 \\ 
  FB & 0.003 & \textbf{0.983} & 0.001 & 0.005 & 0.005 & 0.004 \\ 
  TSLA & 0.068 & 0.040 & 0.009 & \textbf{0.824} & 0.038 & 0.021 \\ 
  BRK.B & 0.000 & 0.000 & \textbf{1} & 0.000 & 0.000 & 0.000 \\ 
  V & 0.006 & 0.019 & 0.006 & 0.025 & \textbf{0.933} & 0.010 \\ 
  JNJ & 0.004 & 0.013 & 0.004 & 0.013 & \textbf{0.959} & 0.008 \\ 
  WMT & 0.032 & 0.045 & 0.022 & 0.032 & 0.035 & \textbf{0.834} \\ 
  JPM & 0.096 & 0.067 & 0.055 & 0.122 & 0.089 & \textbf{0.572} \\ 
  MA & 0.129 & 0.160 & 0.017 & \textbf{0.499} & 0.087 & 0.109 \\ 
  PG & 0.087 & 0.071 & 0.120 & 0.117 & 0.118 & \textbf{0.486} \\ 
  UNH & 0.010 & \textbf{0.905} & 0.004 & 0.026 & 0.037 & 0.018 \\ 
  DIS & 0.057 & 0.120 & 0.014 & \textbf{0.619} & 0.128 & 0.061 \\ 
  NVDA & 0.044 & 0.038 & 0.013 & \textbf{0.802} & 0.077 & 0.025 \\ 
  HD & 0.035 & 0.052 & 0.015 & 0.032 & 0.030 & \textbf{0.836} \\ 
  PYPL & 0.189 & \textbf{0.420} & 0.019 & 0.172 & 0.071 & 0.129 \\ 
  BAC & 0.035 & 0.042 & 0.011 & 0.041 & 0.029 & \textbf{0.841} \\ 
   \hline
\end{tabular}
\caption{Membership degrees for top 20 companies in the S\&P 500 index by considering the QCD-FCMn model and a 6-cluster partition.}
\label{tablefuzzy}
\end{table}

On balance, the clustering partition obtained with the QCD-FCMn model is quite consistent with the distances between the points in Figure \ref{mdstop20}.  Cluster $C_4$ contains the companies Apple (AAPL) and Microsoft (MSFT), Tesla (TSLA), MasterCard (MA), Walt Disney (DIS) and NVIDIA (NVDA) with high membership. Cluster $C_1$ consists of Amazon (AMZN) and the two branches of Google (GOOGL, GOOG). It is worth noting that AAPL and MSFT also possess a non negligible membership degree in this cluster. The remaining one of the biggest five companies, Facebook (FB), presents a large membership degree in cluster $C_2$ (0.983). It is noticeable that FB is barely present in clusters $C_1$ and $C_4$. This fact suggests that its business model greatly differs from that of its tech giants counterparts. Cluster $C_2$ is also constituted by the health insurance company UnitedHelath (UNH) with high membership and by the online payments company PayPal (PYPL). Interestingly enough, this latter company is the more spread out among the six clusters. Indeed, most of the companies in Table \ref{tablefuzzy} offer PYPL as a form of online payment in their corresponding websites. Thus, it is expected that the financial behaviour of PYPL is somehow related to that of the remaining corporations. The multinational conglomerate holding company Berkshire Hathaway (BRK.B) forms itself an isolated cluster ($C_3$), which is congruent with  Figure \ref{mdstop20} where BRK.B is represented as the most outlying point. An economical explanation of this fact can be easily obtained: BRK.B is not the typical corporation. This business model consists of investing in and holding dozens of major public and private companies. Thus, it is no surprising that it shares no similarity with the rest of the firms. Cluster $C_5$ if formed by Visa (V) and Johnson \& Johnson (JNJ) presenting large membership degrees. Finally, cluster $C_6$ includes the three consumer-based corporations Walmart (WMT), Procter \& Gamble (PG) and Home Depot (HD), and the two banking companies JPMorgan Chase (JPM) and Bank of America (BAC). 

Note that a great deal of information can be extracted from the resulting partition by virtue of the fuzzy nature of the clustering partition. For instance, by looking at  Table \ref{tablefuzzy}, one investor could determine that, whereas the financial behaviour of BRK.B is not shared by any of the remaining firms, there are another companies as PYPL and PG which show a changing behaviour. This type of insights could be invaluable in order to make informed investing decisions.

\subsection{Fuzzy clustering of air pollution data}\label{subsectionairpollution}

Now we develop a study case related to the non-supervised classification of geographical zones in terms of their temporal records of air pollutants. In the Spanish autonomous community of Galicia the air quality is analysed by a monitoring network consisting of 14 public and 33 private stations situated at different locations. These 47 stations provide hourly data on air pollutant concentration. The following pollutant concentrations: SO$_2$, NO, NO$_2$, NO$_\text{X}$, CO, O$_3$, PM$_{10}$ (particulate matter 10 micrometers or less in diameter) and PM$_{2,5}$ (particulate matter 2.5 micrometers or less in diameter) are recorded in at least one station of the network. 

We considered trivariate time series of hourly concentrations of nitrogen dioxide (NO$_2$), ozone (O$_3$) and nitrogen monoxide (NO) during the whole year 2018 in 20 different stations. The choice of this subset of pollutants was based on (1) several studies have uncovered serious health effects associated with the continuous exposure to high levels of NO$_2$, O$_3$ and NO, \cite{elsayed1994toxicity, mehlman1987toxicity, depayras2018hidden} and (2) they were the most monitored gases throughout the network. It is important to highlight that our intention is only to show the usefulness of the proposed clustering algorithms without seeking to give any type of environmental implications, although this study together with other analyses could lead to the taking of some steps in order to reduce pollution. The 20 corresponding stations are Ferrol-Parque RS (FE), Coruña-Torre de Hércules (CO-T), Coruña-Riazor (CO-R), Lugo-Fingoy (LU), Santiago-Campus (SDC-C), Santiago-San Caetano (SDC-SC), Sur (SU), Pontevedra-Campolongo (PO-CL), Vigo-Coia (VGO-CO), Vigo-Lope de Vega (VGO-L), Ponteareas (PT), Ourense-Gómez Franqueira (OR), Pontevedra-Campelo (PO-CP), Fraga Redonda (FR), Xove (XO), Vigo-Citroën (VGO-CT), Paiosaco-Laracha (PA), Magdalena (MA), Louseiras (LO) and Mourence (MO). All data were sourced from the website of Ministry for the Ecological Transition and the Demographic challenge\footnote{https://www.miteco.gob.es/es/calidad-y-evaluacion-ambiental/temas/atmosfera-y-calidad-del-aire/calidad-del-aire/evaluacion-datos/datos/Datos\_oficiales\_2018.aspx}. Table \ref{tabletop20stations} contains some general information about the location of the stations. The categorization of the location of each station as ``urban'', ``suburban'', ``rural'' and ``near power plant'' was made based on information provided in the above website. Thus, from an environmental point of view, it is reasonable to think that the joint behaviour of the concentration of the three considered gases is different depending on where the station is situated. 

\begin{table}
	\centering
	\begin{tabular}{ccc} \hline 
		Station                  & Abbreviation & Location area  \\ \hline   
		Ferrol-Parque RS         & FE           & Suburban         \\
		Coruña-Torre de Hércules & CO-T         & Urban            \\
		Coruña-Riazor            & CO-R         & Urban            \\
		Lugo-Fingoy              & LU           & Urban            \\
		Santiago-Campus          & SDC-C        & Suburban         \\
		Santiago-San Caetano     & SDC-SC       & Urban            \\
		Sur                      & SU           & Rural            \\
		Pontevedra-Campolongo    & PO-CL        & Urban            \\
		Vigo-Coia                & VGO-CO       & Urban            \\
		Vigo-Lope de Vega        & VGO-L        & Urban            \\
		Ponteareas               & PT           & Suburban         \\
		Ourense-Gómez Franqueira & OR           & Urban            \\
		Pontevedra-Campelo       & PO-CP        & Rural            \\
		Fraga Redonda            & FR           & Near power plant \\
		Xove                     & XO           & Rural            \\
		Vigo-Citroën             & VGO-CT       & Urban            \\
		Paiosaco-Laracha         & PA           & Near power plant            \\
		Magdalena                & MA           & Near power plant \\
		Louseiras                & LO           & Near power plant \\
		Mourence                 & MO           & Near power plant \\ \hline 
	\end{tabular}
\caption{Location information of the 20 stations monitoring the air quality in the community of Galicia.}
\label{tabletop20stations}
\end{table}

\begin{figure}
	\centering
	\includegraphics[width=1\textwidth]{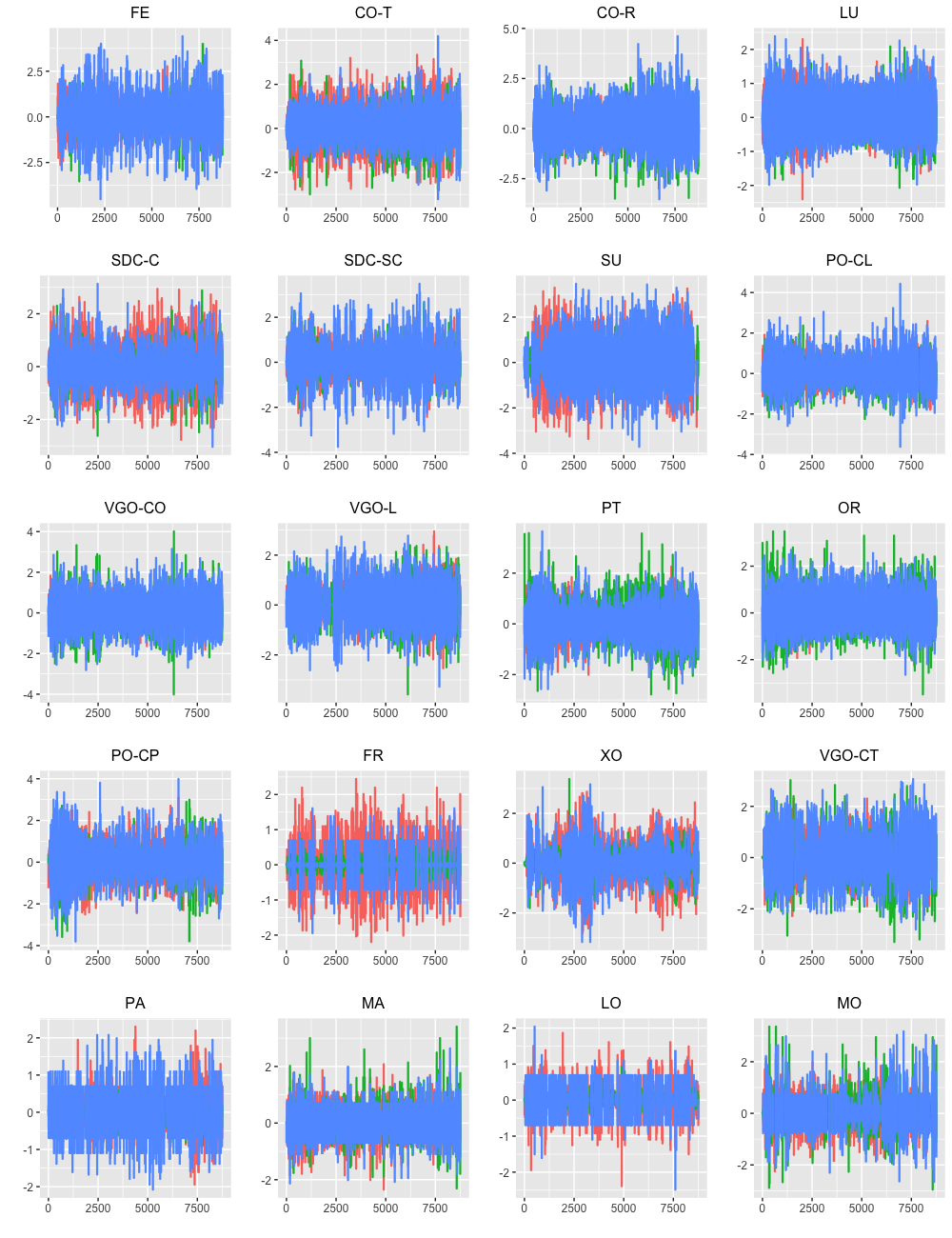}
	\caption{Transformed levels of NO$_2$ (red colour) O$_3$ (green colour) and NO (blue colour) in the 20 monitoring stations of Galicia.}
	\label{top20stations}
\end{figure}

The 20 MTS available are formed by $T=8760$ hourly records and are non-stationary in mean. For this reason, the former series were transformed by taking the first differences of the natural logarithm of the original values. The new series are depicted in Figure \ref{top20stations}. It can be observed that the behaviour of the trivariate series is substantially different among the considered stations. On the other hand, it is reasonable to think that a fuzzy behaviour might be present, with MTS sharing features of distinct and well-defined patterns of hourly changes of concentrations of NO$_2$, O$_3$ and NO.

\begin{figure}
	\centering
	\includegraphics[width=1\textwidth]{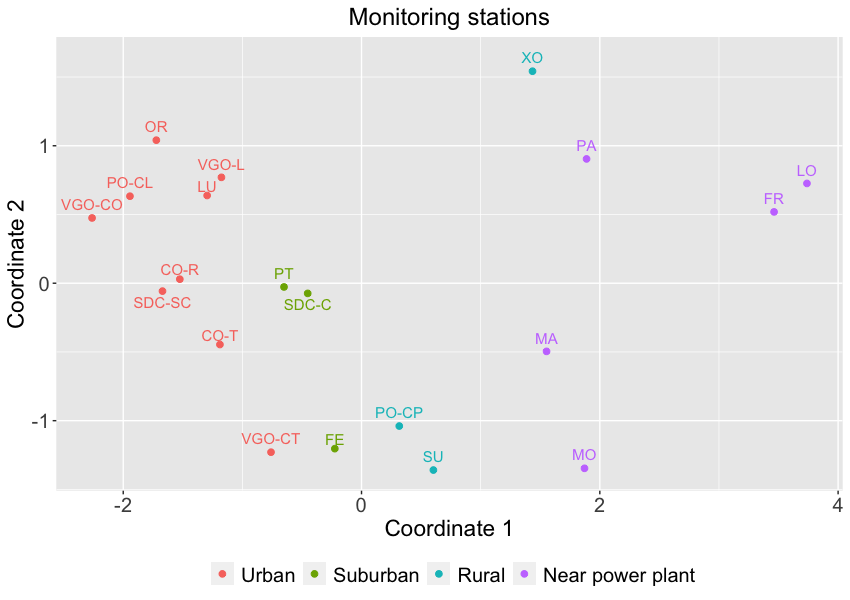}
	\caption{Two-dimensional scaling plane based on the QCD distances for the hourly levels of NO$_2$, O$_3$ and NO in the 20 monitoring stations of Galicia.}
	\label{mdstop20stations}
\end{figure}

First, as in the above case study, we carried out a 2DS based on the QCD-dissimilarity matrix. The resulting 2DS plane in Figure \ref{mdstop20stations}, whose respective R-squared value is 0.8921, gives illustrative insights into the proximity of the time series according to the QCD-based distance. The points have been coloured according to the categories introduced in Table \ref{tabletop20stations} concerning the location of the stations.

Figure \ref{mdstop20stations} shows that the QCD-based distance effectively captures the underlying categories of the stations according to Table \ref{tabletop20stations}. In particular the first coordinate clearly separates the stations with regards to the specific location. Stations situated in urban areas are located in the left side, whereas those positioned near a power plant are placed in the right side. The remaining categories, which correspond to suburban and rural locations, lie somewhere in the middle. It is worth pointing out that the five stations situated near power plants also pertain to rural zones. Note that, whereas the distinction between urban stations and those located in rural regions or near a power plant is evident, the classification of the suburban locations seems vaguer, implying that the dependence relationship of the three pollutants could exhibit traits related to both urban and rural areas. Therefore, a fuzzy clustering approach seems far more suitable to tackle the grouping task of the 20 stations than a hard clustering one. 

By observing Figure \ref{mdstop20stations} and ignoring the underlying categories, one could hypothesize the existence of three or four underlying clusters. There is a first cluster formed by all the urban stations except for VGO-CT together with PT and SDC-C. VGO-CT, FE, PO-CP and SU constitute another cluster, as well as the two stations near a thermal power plant, MA and MO. Finally, XO, PA, FR and LO could be considered to compose one or two different clusters.

As in the preceding application, the QCD-FCMn approach was applied to the series in Figure \ref{top20stations}. Selection of the optimal values for $m$ and $C$ was effectuated by using the same procedure as in the above analysis. After performing the corresponding steps, the existence of three major groups ($C=3$) was concluded, which is consistent with the 2DS plot in Figure \ref{mdstop20stations}. The optimal value for the fuzziness coefficient was $m=1.9$.

The resulting 3-cluster fuzzy partition is displayed in Table \ref{fuzzytablestations}. For a given element, the highest membership degree is shown in bold provided that its value is larger than 0.6. As 3 clusters are being considered, this cutoff seems a sensible choice. When the three membership values are below this cutoff, the corresponding quantities are written in italic font. Fundamentally, the model QCD-FCMn produces the expected classification by grouping the transformed series according to the kind of location where the respective stations are placed. Regarding the urban locations, they are grouped together in cluster $C_1$ with a high membership aside from VGO-CT, which mainly pertains to cluster $C_3$. This cluster brings together the mentioned VGO-CT, two rural stations (SU and PO-CP) and one suburban (FE) station. The misclassification of VGO-CT with respect to the underlying location can be explained because, despite being in the same city as VGO-CO and VGO-L, it is located in a car factory, probably exposing himself to special types of emissions avoided by VGO-CO and VGO-L. The last cluster, $C_2$, includes three of the stations close to a thermal power plant. 

\begin{table}
\centering
\begin{tabular}{ccccc}
  \hline
Station  & Area & $C_1$ & $C_2$ & $C_3$ \\ 
  \hline
FE & Suburban & 0.113 & 0.035 & \textbf{0.853} \\ 
  CO-T & Urban & \textbf{0.701} & 0.041 & 0.258 \\ 
  CO-R & Urban & \textbf{0.799} & 0.042 & 0.159 \\ 
  LU & Urban & \textbf{0.952} & 0.010 & 0.039 \\ 
  SDC-C & Suburban & \textit{0.452} & \textit{0.044} & \textit{0.504} \\ 
  SDC-SC & Urban & \textbf{0.928} & 0.012 & 0.060 \\ 
  SU & Rural & 0.089 & 0.076 & \textbf{0.835} \\ 
  PO-CL & Urban & \textbf{0.963} & 0.008 & 0.029 \\ 
  VGO-CO & Urban & \textbf{0.938} & 0.014 & 0.048 \\ 
  VGO-L & Urban & \textbf{0.903} & 0.020 & 0.076 \\ 
  PT & Suburban & \textit{0.469} & \textit{0.069} & \textit{0.462} \\ 
  OR & Urban & \textbf{0.944} & 0.014 & 0.042 \\ 
  PO-CP & Rural & 0.022 & 0.012 & \textbf{0.966}\\ 
  FR & Near power plant & 0.018 & \textbf{0.947} & 0.036 \\ 
  XO & Rural & \textit{0.191 }& \textit{0.538} & \textit{0.271} \\ 
  VGO-CT & Urban & 0.227 & 0.043 & \textbf{0.730} \\ 
  PA & Near power plant & 0.049 & \textbf{0.850} & 0.101 \\ 
  MA & Near power plant & \textit{0.088} & \textit{0.345} & \textit{0.567 }\\ 
  LO & Near power plant & 0.023 & \textbf{0.933 }& 0.044 \\ 
  MO & Near power plant & \textit{0.115}& \textit{0.294} & \textit{0.590} \\ 
   \hline
\end{tabular}
\caption{Membership degrees for the 20 monitoring stations in Galicia by considering the QCD-FCMn model and a 3-cluster partition.}
\label{fuzzytablestations}
\end{table}

The stations concerning the fuzziest allocations were SDC-C, PT, XO, MA and MO. SDC-C and PT correspond to suburban areas, and are placed in $C_1$ (urban cluster) with membership values of 0.452 and 0.469, respectively, and in $C_3$ (rural cluster) with membership values of 0.504 and 0.462, respectively. XO is the station displaying the most spread between clusters, corresponding its largest membership degree to cluster $C_2$. Finally, both MA and MO are mainly assigned to clusters $C_2$ and $C_3$, showing a higher membership in the later. Note that all the previous assignments are coherent according to the particularities of each cluster. The stations placed in suburban areas are expected to share traits of both urban and rural regions. In the same way, as the five stations located close to a power plant belong to rural zones, they are presumed to show features typical of rural locations. These results stress the power of a fuzzy allocation  when there exist overlapping classes.

For the sake of illustration and comparison purposes, we have obtained the clustering solutions according to the approaches W-FCMn, C-FCMn and F-FCMn. The selection of $C$ and $m$ was carried out similarly than in the above analyses. The corresponding fuzzy partitions were obtained for the optimal values of $m$ and $C$. Assuming that the true partition is given by the categories standing for the location of each station, the FARI was obtained for the three mentioned procedures as well as QCD-FCMn. Note that this quantity is a good indicator of to what extent each procedure is capable of deriving the underlying geographical distribution. The results are displayed in Table \ref{faristations}. QCD-FCMn significantly outperformed the remaining approaches. Whereas W-FCMn and C-FCMn were able to figure out some insights into the region of each stations, F-FCMn produced a random partition according to this criterion. Lastly, it is worth enhancing that all the procedures have determined some series showing a fuzzy nature, which supports the helpfulness of the fuzzy approach.

\begin{table}
	\centering
	\begin{tabular}{ccccc} \hline 
		Method & QCD-FCMn & W-FCMn & C-FCMn & F-FCMn \\ \hline 
		FARI   & \textbf{0.3649}   & 0.2253 & 0.2083 & 0.0040 \\ \hline 
	\end{tabular}
\caption{FARI obtained by the fuzzy $C$-means procedures in grouping the 20 monitoring stations in Galicia. The ground truth is given by the location of the stations.}
\label{faristations}
\end{table}

\section{Concluding remarks and future work}\label{sectionconcludingremarks}
 
 In this work we have proposed two novel approaches for fuzzy clustering of MTS based on the quantile cross-spectral density (QCD) and principal component analysis (PCA), the so-called QCD-FCMn and QCD-FCMd. The former builds on the traditional fuzzy $C$-means algorithm while the latter employs the fuzzy $C$-medoids. The methods utilise a slight modification of the distance $d_{QCD}$ in our previous work \cite{oriona2020} regarding the projection of the 
 QCD-based features onto the principal components space. The advantages of performing dimensionality reduction via PCA in terms of clustering effectiveness have been shown through a motivating example.
 
 
 To evaluate the performance of QCD-FCMn and QCD-FCMd, we have carried out numerical experiments including scenarios formed by MTS pertaining to well-defined clusters and scenarios involving series equidistant from two clusters. Several types of generating processes were considered. The assessment task was executed in two different fashions. Concerning scenarios lacking a switching series, a fuzzy extension of the Adjusted Rand Index was considered. This way, the quality of the resulting fuzzy partition, in the sense of assigning high membership values to the correct clusters, was directly evaluated. On the other hand, the capability of the techniques in scenarios incorporating an equidistant series was measured also by taking into account their ability to determine the fuzzy nature of this series. The methods were compared with other alternative dissimilarities suggested in the literature. Regardless of the considered models and assessment schemes, QCD-FCMn and QCD-FCMd produced the best results, the former slightly outperforming the latter overall. Both methods inherit the powerful characteristics of QCD, as no requirements about the existence of moments, robustness to changes in the error distribution, and computational efficiency. They also preserve the properties of the former distance $d_{QCD}$, as being able to uncover any type of disparity in the dependence structure of two MTS. Two specific case studies involving environmental and financial databases have illustrated the usefulness of the proposed techniques.
 
 It is worth pointing out that this paper represents an original extension of \cite{oriona2020} in four different ways. First, this manuscript presents explicitly some useful properties of the distance $d_{QCD}$ indicating that this dissimilarity is able to detect any type of discrepancy in the dependence structure of two generating processes whatever their complexity. Second, this work highlights the power of applying dimensionality reduction techniques to QCD-based features in relation to clustering performance. Third, whereas the simulated scenarios in \cite{oriona2020} did not include nonlinear processes, we have considered them here in order to examine the approaches under a vast assortment of generating patterns. Lastly, our study in \cite{oriona2020} was limited to crisp clustering procedures. In this article we introduced fuzzy clustering strategies, thus combining the versatility of the fuzzy logic by permitting overlapping clusters with the high ability of the QCD-based metric to differentiate between underlying mechanisms. In fact, the superiority showed by the proposed dissimilarity over the alternative metrics in a fuzzy context is substantially greater than that in a crisp framework. In short, this paper contributes to the few works on fuzzy clustering of MTS based on generating processes.

 There are indeed some appealing issues for further research in relation to the use of QCD in soft clustering of MTS. Specifically, this work can be extended in two different ways. On the one hand, it would be interesting to obtain robust versions of QCD-FCMn and QCD-FCMd capable of properly neutralizing the effect of outlying MTS. For instance, we could consider the techniques used in \cite{d2014robust} and \cite{lafuente2018robust}, namely the metric approach (by smoothing the distance), the trimmed approach (by trimming away a small proportion of the series) and the noise approach (by considering a noise cluster expected to contain the outlying series). On the other hand, note that, by using QCD to describe an MTS dataset, each MTS is characterized by a set of curves of the form
  \begin{equation}
 \left\{ W \left( \hat{G}^{j_1,j_2}_{T,R} (\omega, \tau,\tau^{\prime}) \right), 1\le j_1, j_2 \le d, \tau, \tau^{\prime} \in \mathcal{T}\right\}, 
 \end{equation}
  where $W(\cdot)$ is used interchangeably to denote the real part and the imaginary part operator. Our numerical studies have revealed that some of these curves contain far more information than others in terms of the generating process of each MTS. Thus, it would be reasonable to create a fuzzy clustering algorithm giving more importance to the functions with more discriminative power. This could be naturally accomplished by introducing weights in the objective functions \eqref{qcd_means} and \eqref{qcd_med}. Even an approach considering only two weights, for real and imaginary parts, respectively, could be devised. The mentioned topics for further research will be properly addressed in the upcoming months.

 \section*{Appendix}
 
 We now derive Properties 1 and 2 in Section \ref{subsectionpropertiesdqcd}. Property 1 follows directly from the definition of QCD, the fact that the smoothed CCR-periodogram is a consistent estimator of this quantity and the definition of $d_{QCD}$. 
 
 To show Property 2, note that assumption on the continuity of the cumulative probability distribution functions implies that there exists a neighbourhood $V$ of $\mathbb{R}^2$ in which $F_{j_1,j_2,l}^1$ and $F_{j_1,j_2,l}^2$ differ from one another. Let $(c,d) \in V$, and therefore $F_{j_1,j_2,l}^1(c,d)\ne F_{j_1,j_2,l}^2(c,d)$. By virtue of Sklar\textquotesingle s theorem, every cumulative distribution function can be expressed by means of its marginals and a unique copula. Hence, if $C_{j_1, j_2, l}^1$ and $C_{j_1, j_2, l}^2$ denote the copula associated with the pairs $(X_{t, j_1}^1, X_{t+l, j_2}^1)$ and $(X_{t, j_1}^2, X_{t+l, j_2}^2)$, respectively, we can write
  \begin{equation}
 \begin{split}
F_{j_1,j_2,l}^1(x_1, x_2)=C_{j_1, j_2, l}^1(F_{j_1}^1(x_1), F_{j_2}^1(x_2)),\\ 
 F_{j_1,j_2,l}^2(y_1, y_2)=C_{j_1, j_2, l}^2(F_{j_1}^2(y_1), F_{j_2}^2(y_2)),
 \end{split}
 \end{equation}
 for all $x_1, x_2, y_1, y_2 \in \mathbb{R}$. Take now $x_1=y_1=c$, $x_2=y_2=d$. From the fact that $F_{j_1,j_2,l}^1(c,d)\ne F_{j_1,j_2,l}^2(c,d)$ and the equality of the marginal distributions, we have
 \begin{equation}  \label{diff.copula}
 C_{j_1, j_2, l}^1(F_{j_1}^1(c), F_{j_2}^1(d)) \ne C_{j_1, j_2, l}^2(F_{j_1}^1(c), F_{j_2}^1(d)).
 \end{equation}
 
 Now, consider the following relationships
  \begin{equation}\label{integral}
 \begin{split}
 \int_{-\pi}^{\pi}{\mathfrak f}^{1}_{j_1,j_2}(\omega, \tau_1, \tau_2)e^{il\omega}d\omega=C_{j_1, j_2, l}^1(\tau_1, \tau_2)+\tau_1 \tau_2, \\
 \int_{-\pi}^{\pi}{\mathfrak f}^{2}_{j_1,j_2}(\omega, \tau_3, \tau_4)e^{il\omega}d\omega=C_{j_1, j_2, l}^2(\tau_3, \tau_4) +\tau_3 \tau_4,
 \end{split}
 \end{equation}
 where ${\mathfrak f}^{i}_{j_1,j_2}$, is the quantile cross-spectral density for the pair of processes $X_{t,j_1}^i$ and $X_{t,j_2}^i$, $i=1,2$, $\omega$ is an arbitrary frequency, and $(\tau_1, \tau_2)$ and $(\tau_3, \tau_4)$ are arbitrary couples of probability levels. Then, denoting by $\tau=F_{j_1}^1(c)$ and $\tau'=F_{j_2}^1(d)$, \eqref{diff.copula} and \eqref{integral} allow to conclude that
  \begin{equation}\label{qcdxy}
{\mathfrak f}^{1}_{j_1,j_2}(\omega', \tau, \tau') \ne {\mathfrak f}^{2}_{j_1,j_2}(\omega', \tau, \tau')
 \end{equation}
 for some $\omega' \in V_{\omega}\subset \mathbb{R}$, with $V_{\omega}$ a neighbourhood where the functions differ. From \eqref{qcdxy} we know that at least one of the facts $\Re({\mathfrak f}^{1}_{j_1,j_2}(\omega', \tau, \tau'))\ne \Re({\mathfrak f}^{2}_{j_1,j_2}(\omega', \tau, \tau'))$ or $\Im({\mathfrak f}^{1}_{j_1,j_2}(\omega', \tau, \tau'))\ne \Im({\mathfrak f}^{2}_{j_1,j_2}(\omega', \tau, \tau'))$ is true. Assume without loss of generality that $\Re({\mathfrak f}^{1}_{j_1,j_2}(\omega', \tau, \tau'))\ne \Re({\mathfrak f}^{2}_{j_1,j_2}(\omega', \tau, \tau'))$. Now, in order to compute the distance $d_{QCD}$, select $\mathcal{T}$ and $\Omega$ such that $\tau, \tau' \in \mathcal{T}$ and $\omega' \in \Omega$. Let $\bm \Psi^{(1)}$ and $\bm \Psi^{(2)}$ be the feature vectors computed from the realizations $\bm X_t^{(1)}$ and $\bm X_t^{(2)}$, respectively. From the definition of $d_{QCD}$, it necessarily exists an integer $k$ such that the  $k$-th components of vectors $\bm \Psi^{(1)}$ and $\bm \Psi^{(2)}$ are given by $\Re(\hat{G}^{1, j_1,j_2}_{T,R}(\omega', \tau, \tau'))$ and $\Re(\hat{G}^{2,j_1,j_2}_{T,R}(\omega', \tau, \tau'))$, respectively, being $\hat{G}^{1,j_1,j_2}_{T,R}$ and $\hat{G}^{2,j_1,j_2}_{T,R}$ the smoothed CCR-periodograms computed from the realizations $\bm X_t^{(1)}$ and $\bm X_t^{(2)}$, respectively. As the smoothed CCR-periodogram is a consistent estimate of the quantile cross-spectral density, consistency also holds for real and imaginary parts so we have
 \begin{equation}
 \begin{split}
 \Re(\hat{G}^{1,j_1,j_2}_{T,R}(\omega', \tau, \tau'))\xrightarrow[p]{}\Re({\mathfrak f}^{1}_{j_1,j_2}(\omega', \tau, \tau')), \\
 \Re(\hat{G}^{2,j_1,j_2}_{T,R}(\omega', \tau, \tau'))\xrightarrow[p]{}\Re({\mathfrak f}^{2}_{j_1,j_2}(\omega', \tau, \tau')),
 \end{split}
 \end{equation}
 from which the convergence in probability of the distance $d_{QCD}$ to some $a \ne 0$ is trivially derived. Note that by choosing another pair $(c',d')$, with $c' \ne c$ and $d' \ne d$ in the neighbourhood $V$, a different pair of probability levels such that the result holds could be extracted. The same is true for another $\omega'' \in V_{\omega}$ This process could be repeated infinitely.

 \section*{Declaration}

 \subsection*{Funding}
 The research of Ángel López-Oriona and José. A. Vilar has been supported by the Ministerio de Economía y Competitividad (MINECO) grant MTM2017-87197-C3-1-P,  the Xunta de Galicia through the ERDF (Grupos de Referencia Competitiva ED431C-2016-015), and the Centro de Investigación de Galicia ``CITIC'', funded by Xunta de Galicia and the European Union (European Regional Development Fund- Galicia 2014-2020 Program), by grant ED431G 2019/01.
 
 \subsection*{Competing interests}
 
 The authors have no conflicts of interest to declare.

 \subsection*{Availability of data and material}
 
 Not applicable.
 
 \subsection*{Code availability}
 
 All the code used for achieving the results presented throughout the paper is available under request.
 
 \subsection*{Authors' contributions}
 
 Not applicable.
 
 \subsection*{}
 
 \subsection*{}

\bibliography{mybibfile}

\end{document}